# Non-conforming multipatches for NURBS-based finite element analysis of higher-order phase-field models for brittle fracture

Khuong D. Nguyen[a,b,c], Charles E.Augarde[d], William M. Coombs[d], H. Nguyen-Xuan[e*], M. Abdel-Wahab[f,g*]

[a]*Department of Engineering Mechanics, Faculty of Applied Science, Ho Chi Minh City University of Technology (HCMUT), Ho Chi Minh City, Vietnam*
[b]*Vietnam National University, Ho Chi Minh City, Vietnam*
[c]*Department of Electrical energy, metals, mechanical constructions and systems, Faculty of Engineering and Architecture, Ghent University, Ghent, Belgium*
[d]*Department of Engineering, Durham University, South Road, Durham, DH1 3LE, UK*
[e]*CIRTech Institute, Ho Chi Minh City University of Technology (HUTECH), Ho Chi Minh City, Vietnam*
[f]*Institute of Research and Development, Duy Tan University, 03 Quang Trung, Da Nang, Vietnam*
[g]*Soete Laboratory, Faculty of Engineering and Architecture, Ghent University, Belgium*

∗ Corresponding author
*Email addresses:* ngx.hung@hutech.edu.vn (H. Nguyen-Xuan); magdabdelwahab@duytan.edu.vn; magd.abdelwahab@ugent.be (M. Abdel-Wahab)

## Abstract

This paper proposes an effective computational tool for brittle crack propagation problems based on a combination of a higher-order phase-field model and a non-conforming mesh using a NURBS-based isogeometric approach. This combination, as demonstrated in this paper, is of great benefit in reducing the computational cost of using a local refinement mesh and a higher-order phase-field, which needs higher derivatives of basis functions. Compared with other approaches using a local refinement mesh, the Virtual Uncommon-Knot-Inserted Master-Slave (VUKIMS) method presented here is not only simple to implement but can also reduce the variable numbers. VUKIMS is an outstanding choice in order to establish a local refinement mesh, i.e. a non-conforming mesh, in a multi-patch problem. A phase-field model is an efficient approach for various complicated crack patterns, including those with or without an initial crack path, curved cracks, crack coalescence, and crack propagation through holes. The paper demonstrates that cubic NURBS elements are ideal for balancing the computational cost and the accuracy because they can produce accurate solutions by utilising a lower degree of freedom number than an extremely fine mesh of first-order B-spline elements.

## 1. Introduction

In computational solid mechanics, the accurate prediction of crack formation and propagation remains a crucial challenge applicable in many areas of engineering practice. Over the years, there has been a wide range of studies undertaken to predict fracture failure mechanisms. While the pioneering work by Griffith [1] laid the foundation for modern fracture mechanics, his method itself cannot determine the shape of the crack pattern, or predict crack branching and interaction. Similarly, the concept of the stress intensity factor (SIF), first proposed by Irwin [2], determines the intensity of the stress in the zone nearly by the crack tip, however, the energy of the whole domain is not involved. This can be overcome by several methods based on minimised energy as set out in the works by Francfort [3], Buliga [4], Dal Maso [5], and Bourdin [6].

A crack is a discontinuous zone in an otherwise continuous material, and both *discrete* and *smeared* approaches have been proposed to model a crack pattern using finite elements. In the discrete methods, a discontinuity is created within elements by an enriched formulation of displacement variables using a partition of unity methods (PUMs). The extended finite element method (XFEM) [7] is an outstanding candidate; however, it has some drawbacks. For instance, it not only is a criterion for crack propagation needed, but it also faces difficulties in the cases of crack merging and branching between multiple cracks, especially in three-dimensional fracture. Rabczuk [8] has proposed a simplified meshfree method, such as cracking particles, to improve some drawback of XFEM. It can treat complex patterns involving crack branching and crossing and the nucleation of cracks, but its limitation is less accurate than the others. A combination of the screened Poisson equation and local mesh refinement is proposed by Areias and co-workers [9, 10] to compute damage and fracture problem on both 2D and 3D crack propagations. On the other hand, smeared methods attempt to avoid these issues by using a scalar auxiliary variable to model the appearance and vicinity of a sharp crack surface. This scalar is often termed a *phase-field* variable which has values in both the continuous and discontinuous zones. Miehe [11] first proposed such a phase-field model for modelling complicated fracture patterns, including multiple cracks, crack merging, branching, kinking and nucleation without using any criterion and coupling with finite element analysis as a multi-field problem. Some modifications of the basic phase-field formulation are proposed by Ambati [12] in order to decrease computational time, and phase-field brittle fracture has been extended to ductile fracture models in several papers [13-15]. In addition, the approach has been validated and demonstrated on several complicated fracture problems fatigue problems [16], thermo-elastic solids [17, 18], fluid-saturated porous media [19], piezoelectric and ferroelectric materials [20], microstructures [21] and multi-scale problems [22]. In particular, the phase-field model can be applied successfully to solve crack propagation in complicated material structures, for instance, composite materials [23], fibre-reinforced composite materials [24], functionally graded materials [25] and multi-phase materials [26]. Moreover, the phase-field approach is applied successfully to simulate the crack propagation on the thin shell [27, 28]. These and many other recent studies demonstrate clearly that the phase-field approach has much promising potential a wide scope of applications. However, a key drawback in the standard finite element method-based phase-field model is the need for an extremely fine discretisation (i.e. mesh) in the predicted crack zone in order to capture complex crack patterns represented by the gradients of the phase-field variable. In order to deal with these sometimes excessive computational costs, some techniques applied to other finite element modelling have been proposed, such as adaptive re-meshing [29-32].

A similarly significant advance in computational mechanics in recent years has been the move to close the gap between Computer-Aided Design (CAD) and Finite Element Analysis (FEA) using isogeometric analysis (IGA) as first described by Hughes and co-workers [33]. The key idea is to imply Non-Uniform Rational B-splines (NURBS), widely used to represent geometry in CAD software, as basis functions for the finite element analysis step. The benefits of NURBS basis functions are that they produce the exact conic geometry at a coarse level and therefore re-meshing can be carried out from this level, without further geometry information. They have also been shown to have advantages of flexibility [34] in order to refine and elevate their order. Last but not least, the arbitrary high-order continuity delivered automatically between elements leads to numerous advantages in problems requiring higher derivatives. In the last decade, IGA has been studied and applied widely across computational mechanics, including phase-field fracture modelling, and successfully. The high-order of the NURBS basis functions gives a promising approach for higher-order phase-field theory, specified by Borden [35] leading to an improved convergence rate. Usually, a NURBS mesh can be refined to reach an extremely fine level, but this is *global* refinement, and one cannot mesh in a local zone of a multi-patch problem. Very few solutions have to date been proposed for IGA methods applied to computational fracture mechanics which overcome this cost, including the use of T-splines [36, 37] and hierarchical refinement meshes [38]. However, although they can help to reduce the number of variables, their implementations are very complicated due to the demands of Bézier extraction for NURBS and the need for complex data structures. A non-conforming mesh or local refinement of a multi-patch problem may be the solution to overcome this issue. There are many recent papers on production of a local refinement mesh within a larger coarse discretisation often referred to non-matching approaches, for instance using a penalty formulation [39], Lagrange multiplier method [40], and Nitsche's method [41]. However, each method has drawbacks which will be familiar. Users of the penalty method need to choose a sufficient penalty number to gain a correct solution without leading to ill-conditioning. The Lagrange multiplier method will increase the number of variables while Nitsche's method can be affected by the need for many iterations to establish the coupling between two non-matching meshes.

A recent paper provides a way to achieve the goal of local refinement without some of the drawbacks of existing methods, and it is the approach proposed here for the first time, for phase-field fracture modelling. Virtual Uncommon-Knot-Inserted Master-Slave (VUKIMS) coupling proposed in [42] and is described in more detail below. In this paper, we aim to exploit an efficient computational approach of crack propagation computation in brittle fracture problems by adopting higher-order elements with the NURBS non-matching mesh and higher-order phase-field formulations. VUKIMS is an excellent approach to obtain a local refinement mesh using NURBS geometry in many cases of complex behaviour of crack patterns, such as crack propagation adjacent to a hole or interaction of multiple cracks. No previous studies using these non-matching NURBS mesh methods in order to solve the fracture problem by using phase-fields. High computational cost is addressed here on two fronts, the use of a local refinement mesh and a higher-order phase-field. A variety of numerical examples is considered not merely to verify the solution in comparison with several published studies but also to demonstrate the efficiency of this approach.

## 2. Non-conforming multipatches in a NURBS-based finite element approach

### 2.1. *A brief review of NURBS functions*

A brief review of the non-uniform rational B-spline (NURBS) formulation is presented in this section. More detailed formulations are described in [43]. Firstly, a NURBS surface, $\mathbf{S}(\xi,\eta)$, in case of order *p* in ξ-direction and order *q* in η-direction, can be depicted as

$$\mathbf{S}_{\xi,\eta} = \sum_{i=1}^{n}\sum_{j=1}^{m} R_{i,j}^{p,q}(\xi,\eta) \mathbf{P}_{i,j} \tag{1}$$

where $R_{i,j}^{p,q}(\xi,\eta)$ are the bivariate NURBS basis functions which are determined from the univariate B-spline basis functions, for instance, $N_{i,p}$ and $M_{j,q}$. They are determined on the Ξ and H knot vectors which are in the ξ-direction and η-direction, respectively,

$$R_{i,j}^{p,q}(\xi,\eta) = \frac{N_{i,p}(\xi)M_{j,q}(\eta)w_{i,j}}{\sum_{i=1}^{n}\sum_{j=1}^{m} N_{i,p}(\xi)M_{j,q}(\eta)w_{i,j}} \tag{2}$$

and $\mathbf{P}_{i,j}$ stand for the $n \times m$ control points, and $w_{i,j}$ are the corresponding weights.

The univariate B-spline basis function, $N_{i,p}(\xi)$, can be expressed an open, non-uniform knot vector, $\Xi = \{\xi_1, \xi_2, \ldots, \xi_{n+p+1}\}$ with $\xi_i \leq \xi_{i+1}$ and $i = 1,2,\ldots,n+p$, where n is a control point number, $p$ is the order of a function with non-decreasing sequence numbers in a parametric space of [0,1] by the Cox-de Boor formulation for $p = 1,2,3,\ldots$

$$N_{i,p}(\xi) = \frac{\xi - \xi_i}{\xi_{i+p} - \xi_i} N_{i,p-1}(\xi) + \frac{\xi_{i+p+1} - \xi}{\xi_{i+p+1} - \xi_{i+1}} N_{i+1,p-1}(\xi). \tag{3}$$

In particular, the basis function is described in the case of zero-order, $p = 0$, as

$$N_{i,0}(\xi) = \begin{cases} 1 & \text{if } \xi_i \leq \xi < \xi_{i+1} \\ 0 & \text{otherwise} \end{cases} \tag{4}$$

where, $1 \leq i \leq n$, $p \geq 1$, and $\frac{0}{0}$ is considered as zero. For instance, the quadratic order B-spline basis functions which are constructed from an open knot vector of $\Xi = \{0, 0, 0, 0.25, 0.5, 0.75, 0.75, 1, 1, 1\}$ are plotted in Figure 1.

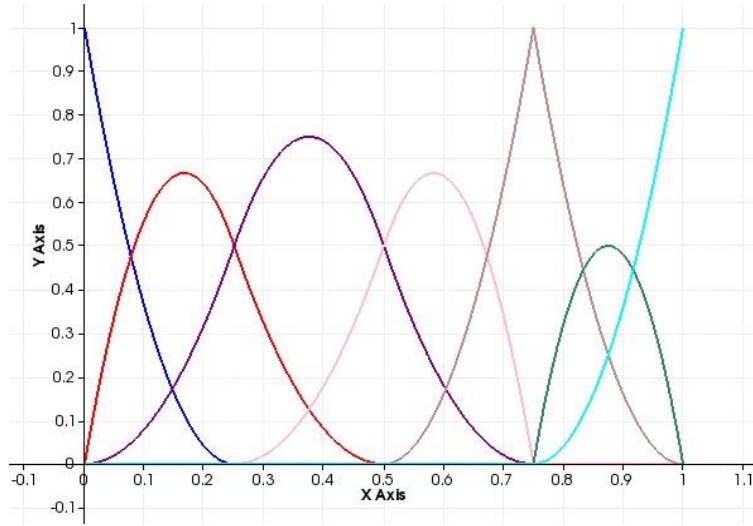

Figure 1. Quadratic basis functions for the knot vector of $\Xi = \{0, 0, 0, 0.25, 0.5, 0.75, 0.75, 1, 1, 1\}$.

A NURBS basis function can be used as a finite element shape function by applying the isoparametric paradigm for phase-field and displacement field approaches. In IGA, the displacement and phase-field variable formulations are described as

$$\mathbf{u}(\xi,\eta) \approx \hat{\mathbf{u}}(\xi,\eta) = \sum_{i=1}^{n}\sum_{j=1}^{m} R_{i,j}^{p,q}(\xi,\eta) \cdot \mathbf{u}_{i,j}, \tag{5}$$

$$\phi(\xi,\eta) \approx \hat{\phi}(\xi,\eta) = \sum_{i=1}^{n}\sum_{j=1}^{m} R_{i,j}^{p,q}(\xi,\eta) \cdot \phi_{i,j} \tag{6}$$

where $\mathbf{u}_{i,j}$ and $\phi_{i,j}$ are displacement and phase-field variables at control point $\mathbf{P}_{i,j}$ on the NURBS surface geometry, respectively.

The first-order derivative of $R_{i,j}^{p,q}(\xi,\eta)$, concerning each parametric coordination, e.g. ξ, is derived from Eq.(2) as

$$\frac{\partial R_{i,j}^{p,q}(\xi,\eta)}{\partial \xi} = \frac{\frac{\partial N_{i,p}(\xi)}{\partial \xi}M_{j,q}(\eta)w_{i,j}W(\xi,\eta) - \frac{\partial W(\xi,\eta)}{\partial \xi}N_{i,p}(\xi)M_{j,q}(\eta)w_{i,j}}{W(\xi,\eta)} \qquad (7)$$

with

$$W(\xi,\eta) = \sum_{i=1}^{n}\sum_{j=1}^{m} N_{i,p}(\xi)M_{j,q}(\eta)w_{i,j}, \qquad (8)$$

$$\frac{\partial W(\xi,\eta)}{\partial \xi} = \sum_{i=1}^{n}\sum_{j=1}^{m} \frac{\partial N_{i,p}(\xi)}{\partial \xi}M_{j,q}(\eta)w_{i,j}. \qquad (9)$$

For more detail, Figure 2 displays the univariate B-spline basis functions tensor product in the case of two knot vectors Ξ= {0, 0, 0, 0, 0.25, 0.5, 0.75, 1, 1, 1, 1} and H = {0, 0, 0, 0.25, 0.5, 0.75, 0.75, 1, 1, 1}.

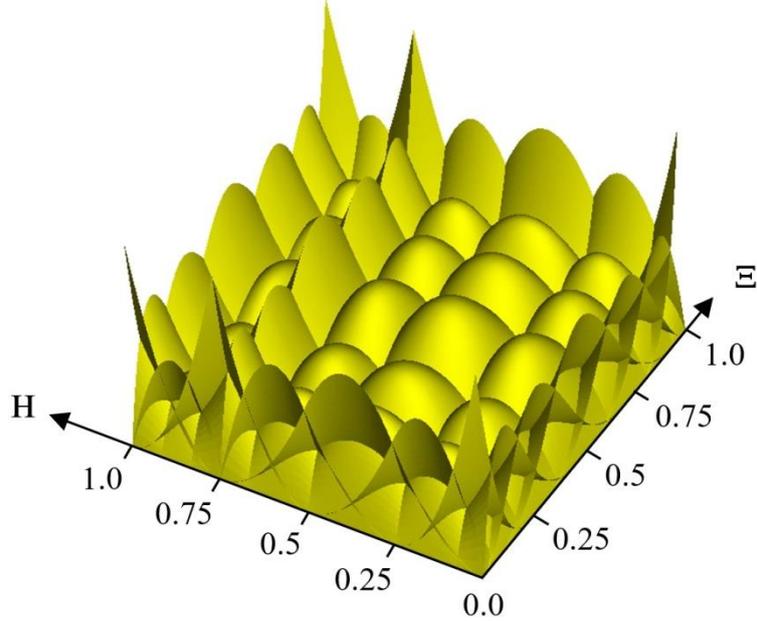

Figure 2. The tensor product of cubic and quadratic basis functions.

### 2.2. Non-conforming multipatch

While the phase-field modelling approximates a crack and its growth, to be successful, the mesh of the region surrounding the crack path needs to be highly refined. In most cases, the proportion of length scale, $l_0$, to the adequate element size, $h$, is chosen as two, which supplies a sufficiently accurate solution without over-resolving the crack [36]. A nonconforming mesh of multiple patches (a "multi-patch") can be applied to deal with this issue, that of providing sufficient refinement for a propagating crack. There are numerous methods for the coupling of nonconforming mesh patches proposed in recent years, for instances, the penalty formulation [39], the mortar method [40, 44], Nitsche's method [41]. However, these methods have some drawbacks that need to be modified [42]. Virtual Uncommon-Knot-Inserted Master-Slave (VUKIMS) coupling proposed by Coox and co-workers [42] is based on master-slave interface constraints and is a robust and straightforward method for a multi-patch model. These master-slave couplings depend on the mesh, which is generated in a geometry-creating step. These couplings are created by establishing a constraint between two interfaces on two patches. Typically, in traditional IGA using NURBS basis functions, two faces must be contained in a two matched control point set. Moreover, if the mesh needs to be refined, all control points must be changed to maintain one-to-one matching of the refined control points. It is a global refinement.

On the other hand, VUKIMS can perform a local refinement using virtual refinement operators. This idea is similar to Bézier extraction operators [45] and gives VUKIMS coupling flexibility and simplicity. This method is valid not only for nonconforming interfaces but also for conforming interfaces. In the conforming patches, the VUKIMS coupling automatically couples one-to-one the interface variables. The VUKIMS patch coupling is established on a mathematical level by a coupling matrix between the master and slave degree of freedoms. In a multi-patch geometry, as shown in Figure 3, the corresponding degree of freedoms (DOFs) is combined for each interface with a master-slave relationship depicted as:

$$\mathbf{d}_s = \mathbf{T}_{sm}\mathbf{d}_m \tag{10}$$

where $T_{sm}$ is the coupling matrix with $n_s \times n_m$ dimension, $n_m$ and $n_s$ are the numbers of control variables $d_m$ and $d_s$ on each interface in the master and slave patches, respectively. Coox et al. proposed that the slave patch is chosen to be finer than the master patch with the same order. This choice keeps the number of unknown independent variables to a minimum.

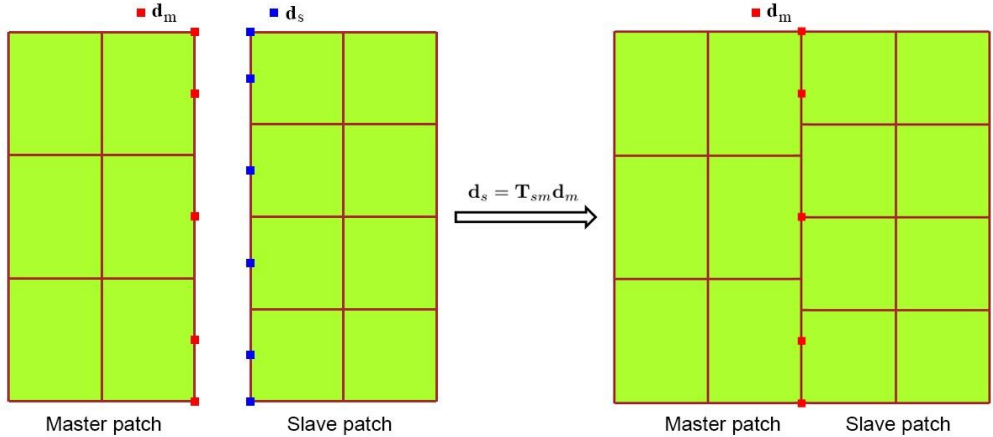

Figure 3. A conceptual illustration of nonconforming patches.

The coupling matrix $T_{sm}$ is described in more detail in [42]. From the relationship in Eq. (10), the coupling matrix can be converted into one linking master and slave DOFs, $\mathbf{u}_M$ and $\mathbf{u}_S$, respectively. These interface constraints are described as follows as

$$\mathbf{u}_S = \mathbf{T}_{SM}\mathbf{u}_M \tag{11}$$

where $T_{SM}$ establishes both dependencies and independencies of all DOFs selected from $T_{sm}$ in Eq. (10). All DOFs can be split into three groups, $\mathbf{u}_O$, $\mathbf{u}_I$ and $\mathbf{u}_D$, representing DOFs belonging to the control points and not belonging to any interfaces, independent and dependent DOFs, respectively. In general, dependent DOFs are slave DOFs while independent DOFs which are master DOFs do not slave any others. The coupling matrix $T_{DI}$ between independent and dependent DOFs is

$$\mathbf{u}_D = \mathbf{T}_{DI}\mathbf{u}_I \tag{12}$$

where $\mathbf{T}_{DI}$ can be extracted from $\mathbf{T}_{SM}$. In general, the global system of equations is expressed as

$$\mathbf{K}\tilde{\mathbf{u}} = \mathbf{f} \tag{13}$$

where $\mathbf{K}$ is the stiffness matrix, $\mathbf{u}$ is the unknown variable vector, and $\mathbf{f}$ is the force vector. Eq. (13) can be rewritten in terms of the three groups of unknown variables as follows

$$\mathbf{K} = \begin{bmatrix} \mathbf{K}_{OO} & \mathbf{K}_{OI} & \mathbf{K}_{OD} \\ \mathbf{K}_{OI}^T & \mathbf{K}_{II} & \mathbf{K}_{ID} \\ \mathbf{K}_{OD}^T & \mathbf{K}_{ID}^T & \mathbf{K}_{DD} \end{bmatrix}, \tilde{\mathbf{u}} = \begin{Bmatrix} \mathbf{u}_O \\ \mathbf{u}_I \\ \mathbf{u}_D \end{Bmatrix}, \mathbf{f} = \begin{Bmatrix} \mathbf{f}_O \\ \mathbf{f}_I \\ \mathbf{f}_D \end{Bmatrix}. \tag{14}$$

Using Eq. (12), Eq. (13) can be transformed to $\mathbf{K}'\tilde{\mathbf{u}}' = \mathbf{f}'$, where

$$\begin{aligned}
\mathbf{K}' &= \begin{bmatrix} \mathbf{K}_{OO} & \mathbf{K}_{OI} + \mathbf{K}_{OD}\mathbf{T}_{DI} \\ (\mathbf{K}_{OI} + \mathbf{K}_{OD}\mathbf{T}_{DI})^T & \mathbf{K}_{II} + \mathbf{K}_{ID}\mathbf{T}_{DI} + \mathbf{T}_{DI}^T(\mathbf{K}_{ID}^T + \mathbf{K}_{DD}\mathbf{T}_{DI}) \end{bmatrix}, \\
\tilde{\mathbf{u}}' &= \begin{Bmatrix} \mathbf{u}_O \\ \mathbf{u}_I \end{Bmatrix}, \\
\mathbf{f}' &= \begin{Bmatrix} \mathbf{f}_O \\ \mathbf{f}_I + \mathbf{T}_{DI}^T \mathbf{f}_D \end{Bmatrix}.
\end{aligned} \tag{15}$$

A reduced system of equations represents the key benefit of using VUKIMS because it can reduce the total number of unknown variables. The dependent DOFs $\mathbf{u}_D$, are determined after independent DOFs $\mathbf{u}_I$, are solved by the reduced system of equations through Eq. (12). Moreover, it is clear to see that by using the VUKIMS, the mesh located at the predicted crack propagation region can be refined locally. The finer mesh must be used locally in the predicted fracture regions, while elsewhere a coarser mesh can be used to save computational time and memory.

3. **Phase-field formulation**

This section presents the formulations of second- and fourth-order phase-field theories in order to describe a brittle fracture of isotropic elastic materials. In this study, the anisotropic formulation of the phase-field method is used for staggered schemes.

*3.1. Phase-field approximation*

Phase-field formulations are proposed by Francfort and Marigo [3], and Bourdin et al. [46] for quasi-static brittle fracture. A sharp crack causing a discontinuous displacement field is replaced by a smeared crack utilising the phase-field variable ($\phi$) as presented in Figure 4. With Griffith's theory, the total energy function suggested for brittle fracture in a variational formulation is a combination of both of elastic energy and fracture energy.

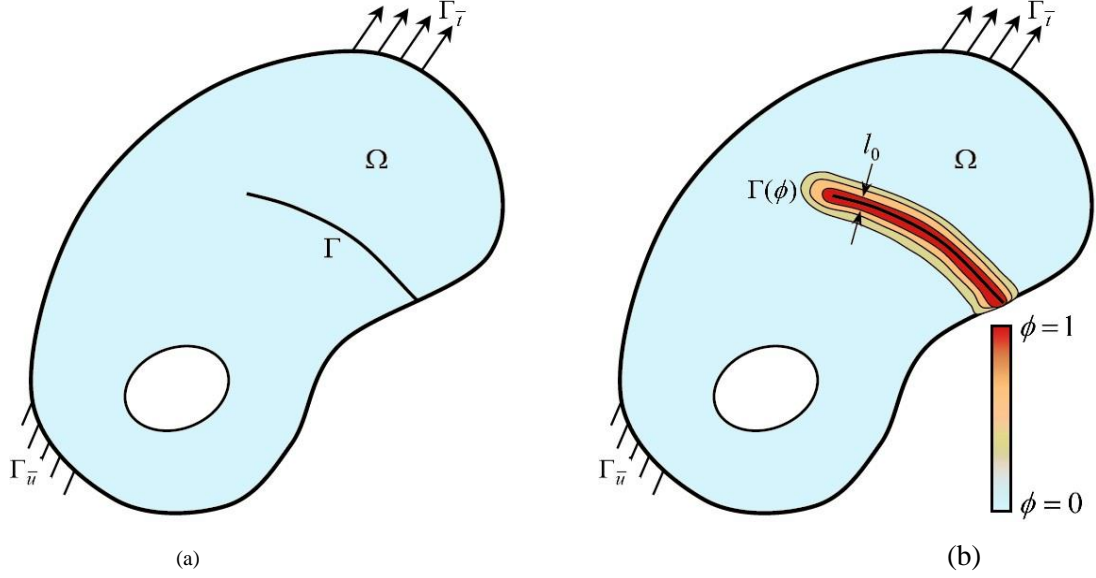

Figure 4. Two ways describe an internal crack: (a) sharp crack and (b) smeared crack.

The energy function is

$$\Pi(\boldsymbol{\varepsilon},\Gamma) = \underbrace{\int_\Omega \psi_{e0}(\boldsymbol{\varepsilon})\mathrm{d}\Omega}_{\text{Elastic energy}} + \underbrace{\int_\Gamma \mathcal{G}_C \mathrm{d}\Gamma}_{\text{Fracture energy}} \qquad (16)$$

where $\mathcal{G}_C$ is the critical energy release density. The infinitesimal strain tensor, which is defined in the case of small strain, is given as

$$\boldsymbol{\varepsilon} = \frac{1}{2}\left(\nabla \mathbf{u} + \nabla^T \mathbf{u}\right). \qquad (17)$$

The undamaged elastic energy, assuming isotropic linear elasticity is given as

$$\psi_{e0}(\boldsymbol{\varepsilon}) = \frac{1}{2}\lambda(tr(\boldsymbol{\varepsilon}))^2 + \mu(\boldsymbol{\varepsilon}:\boldsymbol{\varepsilon}), \qquad (18)$$

in which μ and λ are shear modulus and Lamé's first parameter, respectively. By finding a minimizer of Eq. (16), the propagation of cracks can be predicted as a variational problem. For discrete cracks in the structure, finding the numerical solution of this variational approach can be difficult because of the changing of the crack path in the time-domain requiring re-meshing of the solid domain around the cracks. A scalar-valued phase-field, $\phi \in [0,1]$, is used to approximate a smeared crack in order to solve this issue. Particularly, $\phi = 0$ represents the entire domain, while $\phi = 1$ depicts the fractured domain. For the the fracture energy part, Bourdin et al. [47] introduced a crack density functional, $\psi_{\phi,n}$ to approximate the phase-field

$$\int_\Gamma \mathcal{G}_C \mathrm{d}\Gamma \approx \int_\Omega \mathcal{G}_C \psi_{\phi,n} \mathrm{d}\Omega \qquad (19)$$

where *n* depends on the order of phase-field theory is chosen to approximate phase-field variable *c*. A stress degradation function, $g(\phi)$, proposed to approximate the discontinuity zone, is defined from [11] as $g(\phi) = (1-\phi)^2 + \kappa$. In the case of a fully broken system ($\phi = 1$), we chose a small constant for parameter κ to avoid ill-conditioning. However, Borden [35] proved that this parameter was unnecessary in all calculations, so κ was set zero in this paper. The total energy can be rewritten as

$$\Pi(\mathbf{\varepsilon},\phi) = \int_\Omega g(\phi)\psi_{e0}(\mathbf{\varepsilon})\mathrm{d}\Omega + \int_\Omega \mathcal{G}_C \psi_{\phi,n}\mathrm{d}\Omega. \tag{20}$$

Eq. (18) is an isotropic formulation assumpted asymmetry of the fracture for both of tension and compression parts. Miehe [48] modified the bulk energy density into a distinction between tension and compression parts to represent a decomposition of elastic energy, $\psi_e$, as follows

$$\int_\Omega \psi_e(\mathbf{\varepsilon})\mathrm{d}\Omega = \int_\Omega (g(\phi)\psi_e^+(\mathbf{\varepsilon}) + \psi_e^-(\mathbf{\varepsilon}))\mathrm{d}\Omega \tag{21}$$

where $\psi^+_e$ and $\psi^-_e$ represent the positive and negative components of the strain energies defined as

$$\psi_e^\pm(\mathbf{\varepsilon}) = \frac{\lambda}{2}(\langle\mathrm{tr}(\mathbf{\varepsilon})\rangle^\pm)^2 + \mu\mathrm{tr}[(\mathbf{\varepsilon}^\pm)^2]. \tag{22}$$

The strain tensor is decomposed into positive and negative strain tensors as follows: $\mathbf{\varepsilon} = \mathbf{\varepsilon}^+ + \mathbf{\varepsilon}^-$ and $\mathbf{\varepsilon}^\pm = \sum_{a=1}^{d}\langle\varepsilon_a\rangle^\pm \mathbf{n}_a \otimes \mathbf{n}_a$, where $\varepsilon_a$ and $\mathbf{n}_a$ are the eigenvalues and eigenvectors of the strain tensor, $\mathbf{\varepsilon}$, in the $d$ spatial dimensions, respectively. The Macaulay brackets are defined as: $\langle x \rangle^\pm = \frac{1}{2}(x \pm |x|)$.

Figure 5 illustrates both second-order and fourth-order phase-field approximation of a one-dimensional crack with length-scale $l_0 = 0.1$ as a size damaged zone. These higher-order phase-field theories are depicted in the next section.

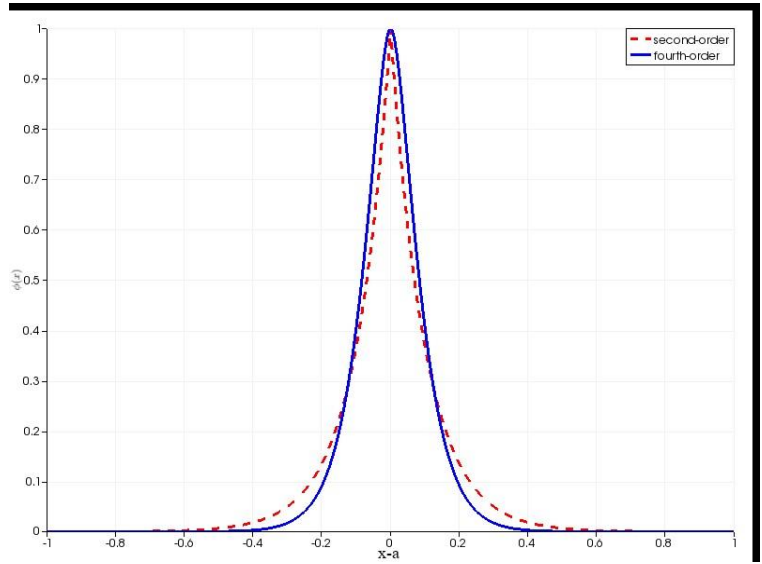

Figure 5. The phase-field approach of the one-dimensional fracture surface.

### 3.1.1. A formulation of a second-order phase-field theory

The fracture energy part of the second-order phase-field theory is given as

$$\int_\Gamma \mathcal{G}_C d\Gamma \approx \int_\Omega \mathcal{G}_C \psi_{\phi,2} d\Omega. \tag{23}$$

The crack density functional of the second-order phase-field theory which has been proposed by Bourdin [46] and utilised by Miehe [11] has the form

$$\psi_{\phi,2} = \frac{\phi^2}{2l_0} + \frac{l_0}{2}(\nabla(\phi))^2. \tag{24}$$

This formulation is denoted as a second-order phase-field theory because Eq. (24) possesses the second-order derivative of the phase-field variable, $\phi$, described as follows

$$\phi(x) = exp(\frac{-|x-a|}{l_0}). \tag{25}$$

*3.1.2. A formulation of the fourth-order phase-field*

In order to describe the crack, Borden [35] introduced a higher-order phase-field formulation which supplied an additional solution regularity. This study proved that this formulation helps to approach not only a better accuracy but also a higher convergence rate for numerical solutions than the second-order phase-field formulation. Hence, the fracture energy part of the fourth-order phase-field theory is determined as

$$\int_\Gamma \mathcal{G}_C d\Gamma \approx \int_\Omega \mathcal{G}_C \psi_{\phi,4} d\Omega. \tag{26}$$

The crack density functional of phase-field variable, $\phi$, is described as

$$\psi_{\phi,4} = \frac{\phi^2}{2l_0} + \frac{l_0}{4}(\nabla(\phi))^2 + \frac{l_0^3}{32}(\Delta\phi)^2 \tag{27}$$

where the phase-field variable, $\phi$, is represented as

$$\phi(x) = \exp\left(\frac{-2|x-a|}{l_0}\right)\left(1 + \frac{2|x-a|}{l_0}\right). \tag{28}$$

*3.2. Governing equations*

Governing equations can be formulated with the constitutive law of the total energy in order to determine the displacement field, **u**, and the phase-field, $\phi$, in a fractured domain outlined Eq. (20). The internal total energy variation is depicted as

$$\delta W_{int} = \delta\Pi(\boldsymbol{\varepsilon},\phi) = \left(\frac{\partial\Pi}{\partial\phi}\right)\delta\phi + \left(\frac{\partial\Pi}{\partial\boldsymbol{\varepsilon}}\right):\delta\boldsymbol{\varepsilon}. \tag{29}$$

Eq. (29) can be derived in two cases of phase-field theories as

- For the second-order formulation:

$$\delta W_{int} = \int_\Omega \frac{\mathcal{G}_C}{l_0} \phi \delta\phi \mathrm{d}\Omega + \int_\Omega \mathcal{G}_C l_0 \nabla\phi \cdot \nabla\delta\phi \mathrm{d}\Omega + \int_\Omega -2(1-\phi)\psi_e^+ \delta\phi \mathrm{d}\Omega + \int_\Omega \boldsymbol{\sigma}\delta\boldsymbol{\varepsilon}\mathrm{d}\Omega \qquad (30)$$

- For the fourth-order formulation:

$$\delta W_{int} = \int_\Omega \frac{\mathcal{G}_C}{l_0} \phi \delta\phi \mathrm{d}\Omega + \int_\Omega \frac{\mathcal{G}_C l_0}{2} \nabla\phi \cdot \nabla\delta\phi \mathrm{d}\Omega + \int_\Omega -2(1-\phi)\psi_e^+ \delta\phi \mathrm{d}\Omega + \int_\Omega \frac{\mathcal{G}_C l_0^3}{16} \Delta\phi \cdot \Delta\delta\phi \mathrm{d}\Omega + \int_\Omega \boldsymbol{\sigma}\delta\boldsymbol{\varepsilon}\mathrm{d}\Omega \quad (31)$$

where the stress tensor is depicted as

$$\begin{aligned}\boldsymbol{\sigma} &= \left[(1-\phi)^2 + \kappa\right]\frac{\partial \psi_e^+}{\partial \boldsymbol{\varepsilon}} + \frac{\partial \psi_e^-}{\partial \boldsymbol{\varepsilon}} = \left[(1-\phi)^2 + \kappa\right]\boldsymbol{\sigma}^+ + \boldsymbol{\sigma}^- \\ &= \left[(1-\phi)^2 + \kappa\right](\lambda\langle\mathrm{tr}(\boldsymbol{\varepsilon})\rangle^+ \mathbf{I} + 2\mu\boldsymbol{\varepsilon}^+) + (\lambda\langle\mathrm{tr}(\boldsymbol{\varepsilon})\rangle^- \mathbf{I} + 2\mu\boldsymbol{\varepsilon}^-)\end{aligned} \qquad (32)$$

where $\boldsymbol{I}$ is the identity tensor.

Furthermore, the external work variation, depending on the external mechanical loading because there is no prescribed external crack phase-field $\phi$ loading, is represented as

$$\delta W_{ext} = \int_\Omega \mathbf{b} \cdot \delta\mathbf{u}\mathrm{d}\Omega + \int_{\partial\Omega_t} \mathbf{t} \cdot \delta\mathbf{u}\mathrm{d}\partial\Omega_t \qquad (33)$$

where $\mathbf{b}$ is a body force and $\mathbf{t}$ is a traction vector which is applied to $\partial\Omega_t$.

For quasi-static process, the equilibrium of the internal and external work increment establishes for deriving the governing equations under a weak form is described as

$$\delta W_{int} - \delta W_{ext} = 0. \qquad (34)$$

By substituting the Eq. (33) combining with Eq. (30) or (31) to Eq. (34), the virtual work statement is given as

- For the second-order formulation:

$$\begin{aligned}&\int_\Omega \frac{\mathcal{G}_C}{l_0} \phi \delta\phi \mathrm{d}\Omega + \int_\Omega \mathcal{G}_C l_0 \nabla\phi \cdot \nabla\delta\phi \mathrm{d}\Omega + \int_\Omega -2(1-\phi)\psi_e^+ \delta\phi \mathrm{d}\Omega \\ &+ \int_\Omega \boldsymbol{\sigma}\delta\boldsymbol{\varepsilon}\mathrm{d}\Omega - \int_\Omega \mathbf{b} \cdot \delta\mathbf{u}\mathrm{d}\Omega - \int_{\partial\Omega_t} \mathbf{t} \cdot \delta\mathbf{u}\mathrm{d}\partial\Omega_t = 0\end{aligned} \qquad (35)$$

- For the fourth-order formulation:

$$\begin{aligned}&\int_\Omega \frac{\mathcal{G}_C}{l_0} \phi \delta\phi \mathrm{d}\Omega + \int_\Omega \frac{\mathcal{G}_C l_0}{2} \nabla\phi \cdot \nabla\delta\phi \mathrm{d}\Omega + \int_\Omega -2(1-\phi)\psi_e^+ \delta\phi \mathrm{d}\Omega + \int_\Omega \frac{\mathcal{G}_C l_0^3}{16} \Delta\phi \cdot \Delta\delta\phi \mathrm{d}\Omega \\ &+ \int_\Omega \boldsymbol{\sigma}\delta\boldsymbol{\varepsilon}\mathrm{d}\Omega - \int_\Omega \mathbf{b} \cdot \delta\mathbf{u}\mathrm{d}\Omega - \int_{\partial\Omega_t} \mathbf{t} \cdot \delta\mathbf{u}\mathrm{d}\partial\Omega_t = 0.\end{aligned} \qquad (36)$$

Gauss theorem is applied to Eq. (35) and (36) becoming as

- For the second-order formulation:

$$\int_\Omega \left\{ -[\nabla \cdot \boldsymbol{\sigma} + \mathbf{b}] \cdot \delta \mathbf{u} + \left[ \mathcal{G}_C \left[ \frac{\phi}{l_0} - l_0 \Delta \phi \right] - 2(1-\phi)\psi_e^+ \right] \delta \phi \right\} d\Omega \qquad (37)$$
$$+ \int_{\partial \Omega_t} [\boldsymbol{\sigma} \cdot \mathbf{n} - \mathbf{t}] \cdot \delta \mathbf{u} d\partial\Omega_t + \int_{\partial\Omega} [\mathcal{G}_C l_0 \nabla \phi \cdot \mathbf{n}] \delta\phi d\partial\Omega = 0$$

- For the fourth-order formulation:

$$\int_\Omega \left\{ -[\nabla \cdot \boldsymbol{\sigma} + \mathbf{b}] \cdot \delta \mathbf{u} + \left[ \mathcal{G}_C \left[ \frac{\phi}{l_0} - \frac{l_0}{2} \Delta \phi + \frac{l_0^3}{16} \Delta(\Delta\phi) \right] - 2(1-\phi)\psi_e^+ \right] \delta \phi \right\} d\Omega \qquad (38)$$
$$+ \int_{\partial \Omega_t} [\boldsymbol{\sigma} \cdot \mathbf{n} - \mathbf{t}] \cdot \delta \mathbf{u} d\partial\Omega_t + \int_{\partial\Omega} \left[ \frac{\mathcal{G}_C l_0}{2} \left[ \nabla\phi - \frac{l_0^2}{8} \nabla(\Delta\phi) \right] \cdot \mathbf{n} \right] \delta\phi d\partial\Omega + \int_{\partial\Omega} \frac{\mathcal{G}_C l_0^3}{16} \Delta\phi \nabla \delta\phi \cdot \mathbf{n} d\partial\Omega = 0$$

where n demonstrates the unit normal vector of the surface $\partial\Omega$. The governing balance equations under a strong form for displacement field can be expressed as

$$\nabla \cdot \boldsymbol{\sigma} + \mathbf{b} = 0 \text{ on } \Omega \qquad (39)$$

with the Neumann-type boundary conditions belong to the displacement field

$$\boldsymbol{\sigma} \cdot \mathbf{n} = \mathbf{t} \text{ on } \partial\Omega_t. \qquad (40)$$

The coupled-fields balance equations under the strong form are presented as

- For the second-order formulation:

$$\mathcal{G}_C \left[ \frac{\phi}{l_0} - l_0 \Delta\phi \right] - 2(1-\phi)\mathcal{H}^+ = 0 \text{ on } \Omega \qquad (41)$$

with the Neumann-type boundary conditions

$$\nabla \phi \cdot \mathbf{n} = 0 \text{ on } \partial\Omega \qquad (42)$$

- For the fourth-order formulation:

$$\mathcal{G}_C \left[ \frac{\phi}{l_0} - \frac{l_0}{2}\Delta\phi + \frac{l_0^3}{16}\Delta(\Delta\phi) \right] - 2(1-\phi)\mathcal{H}^+ = 0 \text{ on } \Omega \qquad (43)$$

with the Neumann-type boundary conditions

$$\left[ \nabla\phi - \frac{l_0^2}{8}\nabla(\Delta\phi) \right] \cdot \mathbf{n} = 0 \text{ and } \Delta\phi = 0 \text{ on } \partial\Omega \qquad (44)$$

where $\mathcal{H}^+ := \max \psi_e^+(\varepsilon)$ is a history-field variable which is a maximum value of positive strain energy [48]. This variable couples weakly the displacement field and the phase field. The history-field variable must be satisfied by the Karush-Kuhn-Tucker conditions for unloading and loading conditions [49] which are depicted as

$$\psi_e^+ - \mathcal{H}^+ \leq 0, \quad \dot{\mathcal{H}} \geq 0, \quad \dot{\mathcal{H}}(\psi_e^+ - \mathcal{H}^+) = 0. \tag{45}$$

*3.3. Variational principles of phase-field formulations*

Using variational principles of the strong form of displacement in Eq. (39) and phase-field in Eq. (41) or (43), the weak form formulations are expressed as

• For the displacement field:

$$\int_\Omega (\boldsymbol{\sigma}\delta\boldsymbol{\varepsilon} - \mathbf{b}\cdot\delta\mathbf{u})\mathrm{d}\Omega - \int_{\partial\Omega_t} \mathbf{t}\cdot\delta\mathbf{u}\,\mathrm{d}\partial\Omega_t = 0 \tag{46}$$

• For the second-order phase-field theory:

$$\int_\Omega \left\{ \mathcal{G}_C\left[\frac{1}{l_0}\phi\delta\phi + l_0\nabla\phi\cdot\nabla\delta\phi\right] - 2(1-\phi)\mathcal{H}^+\delta\phi \right\}\mathrm{d}\Omega = 0 \tag{47}$$

or for the fourth-order phase-field theory:

$$\int_\Omega \left\{ \mathcal{G}_C\left[\frac{1}{l_0}\phi\delta\phi + \frac{l_0}{2}\nabla\phi\cdot\nabla\delta\phi + \frac{l_0^3}{16}\Delta\phi\cdot\Delta\delta\phi\right] - 2(1-\phi)\mathcal{H}^+\delta\phi \right\}\mathrm{d}\Omega = 0. \tag{48}$$

For isogeometric analysis, the displacement field variable, **u**, and the phase-field variable, $\phi$, are approached as

$$\mathbf{u} = \sum_{i=1}^m \mathbf{N}_i^\mathbf{u}\mathbf{u}_i$$
$$\phi = \sum_{i=1}^m N_i\phi_i \tag{49}$$

where $N_i$ is a NURBS basis function which corresponds with control point $i$ of NURBS surface, $m$ is the number of control points per element, and $\mathbf{u_i} = \{u_x, u_y\}^T$ and $\phi_i$ are displacement and phase-field variables of control point $i^{th}$, respectively. Here, a shape function matrix is denoted as

$$\mathbf{N}_i^\mathbf{u} = \begin{bmatrix} N_i & 0 \\ 0 & N_i \end{bmatrix}. \tag{50}$$

The corresponding derivatives can be computed as

$$\boldsymbol{\epsilon} = \sum_{i=1}^m \mathbf{B}_i^\mathbf{u}\mathbf{u}_i, \quad \nabla\phi = \sum_{i=1}^m \mathbf{B}_\mathbf{i}^\phi\phi_i \quad \text{and} \quad \Delta\phi = \sum_{i=1}^m D_i^\phi\phi_i \tag{51}$$

where the strain-displacement matrices are depicted as

$$\mathbf{B}_i^{\mathbf{u}} = \begin{bmatrix} N_{i,x} & 0 \\ 0 & N_{i,y} \\ N_{i,y} & N_{i,x} \end{bmatrix},$$

$$\mathbf{B}_i^{\phi} = \begin{bmatrix} N_{i,x} \\ N_{i,y} \end{bmatrix},$$

$$D_i^{\phi} = N_{i,xx} + N_{i,yy}$$

(52)

where $N_{i,x}$ and $N_{i,y}$ are the first derivatives of the shape function, and $N_{i,xx}$ and $N_{i,yy}$ are the second derivatives of the shape function with respect to $x$ and $y$ directions, respectively. Hence, the variations of both fields and their derivative variables are described as

$$\delta \mathbf{u} = \sum_{i=1}^{m} \mathbf{N}_i^{\mathbf{u}} \delta \mathbf{u}_i, \quad \delta \phi = \sum_{i=1}^{m} N_i \delta \phi_i,$$

$$\delta \boldsymbol{\epsilon} = \sum_{i=1}^{m} \mathbf{B}_i^{\delta \mathbf{u}} \mathbf{u}_i, \quad \nabla \delta \phi = \sum_{i=1}^{m} \mathbf{B}_i^{\phi} \delta \phi_i \quad \text{and} \quad \Delta \delta \phi = \sum_{i=1}^{m} D_i^{\phi} \delta \phi_i.$$

(53)

### 3.4. Staggered scheme algorithm for a phase-field fracture

A staggered solution algorithm which was first proposed by Miehe [48] is the common procedure for crack propagation problems [50-53] to solve the coupled phase-field/elasticity problem. Both fields are solved by minimising the internal potential energy by a Newton-Raphson iteration algorithm with displacement control. The displacement field is solved by the weak form equation which is described in Eq. (46) while the weak form equations of the second- and fourth-order of phase-field theories [35] which are as given in Eq. (47) and Eq. (48), respectively.

The staggered solution scheme for brittle crack propagation in $[t_n, t_{n+1}]$ using a phase-field fracture model is outlined as follows:

(1) Initialisation: The history-field, $\mathcal{H}_n^+$, phase-field, $\phi_n$, and displacement field, $u_n$, at $n^{th}$ loading step are known.

(2) Computing history-field: Updating the maximum history-field depending on displacement $u_n$ belongs to $\mathcal{H}^+ := \max \psi_e^+(\varepsilon)$ and stores them as a history-field value $\mathcal{H}^+$.

(3) The displacement field and phase-field solutions: Updating the current displacement field $\{\mathbf{u}\}_{n+1} = \{\mathbf{u}\}_n + \{\Delta \mathbf{u}\}$ and phase-field $\{\phi\}_{n+1} = \{\phi\}_n + \{\Delta \phi\}$. The increments of the displacement field and phase-field vector, $\{\Delta \mathbf{u} \; \Delta \phi\}^T$, are computed from linear algebraic equations

$$\begin{bmatrix} \mathbf{K}_{ij}^{\mathbf{uu}} & 0 \\ 0 & \mathbf{K}^{\phi\phi} \end{bmatrix} \begin{Bmatrix} \Delta \mathbf{u} \\ \Delta \phi \end{Bmatrix} = \begin{Bmatrix} -\mathbf{r}_i^{\mathbf{u}} \\ -\mathbf{r}_i^{\phi} \end{Bmatrix}$$

(54)

in which the tangent stiffness matrix and the residual of the displacement field are computed as

$$\mathbf{K}_{ij}^{\mathbf{uu}} = \frac{\partial \mathbf{r}_i^{\mathbf{u}}}{\partial \mathbf{u}_j} = \int_{\Omega} (\mathbf{B}_i^{\mathbf{u}})^T \mathbf{C} \mathbf{B}_j^{\mathbf{u}} d\Omega,$$

(55)

$$\mathbf{r}_i^{\mathbf{u}} = \int_{\Omega} (\mathbf{B}_i^{\mathbf{u}})^T \left\{ \left[ (1-\phi)^2 + \kappa \right] \boldsymbol{\sigma}^+ + \boldsymbol{\sigma}^- \right\} d\Omega - \int_{\Omega} (\mathbf{N}_i^{\mathbf{u}})^T \mathbf{b} d\Omega - \int_{\partial \Omega_t} (\mathbf{N}_i^{\mathbf{u}})^T \mathbf{t} d\partial \Omega_t.$$

(56)

The fourth-order tensor C is defined as [51]

$$C_{ijkl} = \frac{\partial \sigma_{ij}}{\partial \varepsilon_{kl}} = \frac{\partial}{\partial \varepsilon_{kl}}\left(\left[(1-\phi)^2 + \kappa\right]\frac{\partial \psi_e^+}{\partial \varepsilon_{ij}} + \frac{\partial \psi_e^-}{\partial \varepsilon_{ij}}\right)$$
$$= \left[(1-\phi)^2 + \kappa\right]\left[\lambda H(\varepsilon_{kk})J_{ijkl} + 2\mu P_{ijkl}^+\right] + \left[\lambda H(-\varepsilon_{kk})J_{ijkl} + 2\mu P_{ijkl}^-\right] \tag{57}$$

where $H(x)$ is the Heaviside function, and a fourth-order tensor, $J_{ijkl}$, is defined as $J_{ijkl} = I_{ij} \otimes I_{kl}$. Projection tensors $P^{\pm}$ are described as [54]

$$\mathbf{P}^{\pm} = \sum_{a=1}^{d}\left(H(\pm\varepsilon_a)\mathbf{Q}_a + \sum_{b=1, b \neq a}^{d} \frac{\langle\varepsilon_a\rangle^{\pm} - \langle\varepsilon_b\rangle^{\pm}}{2(\varepsilon_a - \varepsilon_b)}(\mathbf{G}_{ab} + \mathbf{G}_{ba})\right) \tag{58}$$

with

$$\begin{aligned}\mathbf{M}_a &= \mathbf{n}_a \otimes \mathbf{n}_a, \\ (Q_a)_{ijkl} &= (M_a)_{ij}(M_b)_{kl}, \\ (G_{ab})_{ijkl} &= (M_a)_{ik}(M_b)_{jl} + (M_a)_{il}(M_b)_{jk}. \end{aligned} \tag{59}$$

The tangential stiffness matrix and the phase-field residual vector are computed as
- Second - order theory:

$$\mathbf{K}_{ij}^{\phi\phi} = \frac{\partial r_i^\phi}{\partial \phi_j} = \int_\Omega \left\{ \mathcal{G}_C l_0 (\mathbf{B}_i^\phi)^T \mathbf{B}_j^\phi + \left[2\mathcal{H}^+ + \frac{\mathcal{G}_C}{l_0}\right] N_i N_j \right\} d\Omega, \tag{60}$$

$$r_i^\phi = \int_\Omega \left\{ \mathcal{G}_C \left[\frac{1}{l_0} N_i \phi + l_0 (\mathbf{B}_i^\phi)^T \nabla\phi\right] - 2(1-\phi)\mathcal{H}^+ N_i \right\} d\Omega. \tag{61}$$

- Fourth - order theory:

$$\mathbf{K}_{ij}^{\phi\phi} = \frac{\partial r_i^\phi}{\partial \phi_j} = \int_\Omega \left\{ \frac{\mathcal{G}_C l_0}{2} (\mathbf{B}_i^\phi)^T \mathbf{B}_j^\phi + \left[2\mathcal{H}^+ + \frac{\mathcal{G}_C}{l_0}\right] N_i N_j + \frac{\mathcal{G}_C l_0^3}{16} D_i^\phi D_j^\phi \right\} d\Omega, \tag{62}$$

$$r_i^\phi = \int_\Omega \left\{ \mathcal{G}_C \left[\frac{1}{l_0} N_i \phi + \frac{l_0}{2} (\mathbf{B}_i^\phi)^T \nabla\phi + \frac{l_0^3}{16} D_i^\phi \Delta\phi\right] - 2(1-\phi)\mathcal{H}^+ N_i \right\} d\Omega. \tag{63}$$

Figure 6 shows the staggered scheme algorithm of crack propagation using phase-field theory.

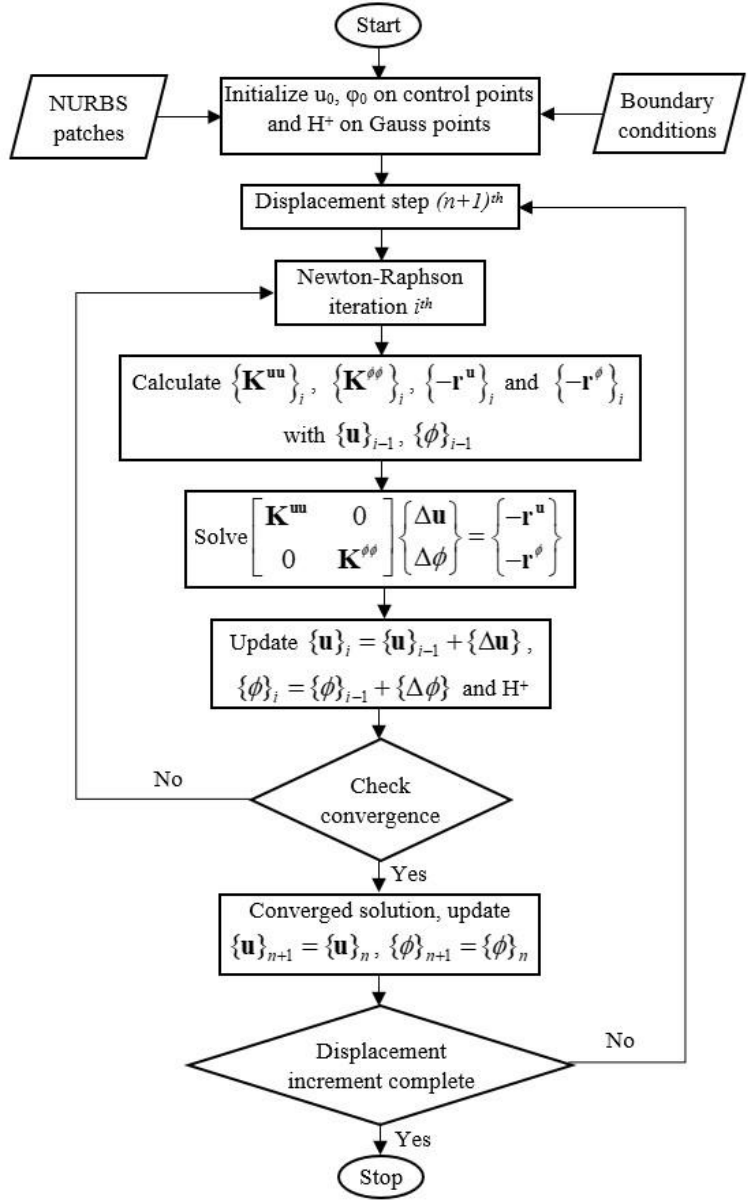

Figure 6. A Flow chart algorithm of the staggered scheme for the phase-field model.

## 4. Numerical examples

In this section, several numerical examples are used to demonstrate the efficiency of the proposed nonconforming mesh IGA approach in modelling crack propagation problems. For all of the numerical examples, the stress state is assumed to be under a plane strain condition, and a Newton-Raphson solution method with displacement control is used to solve the non-linear problems. The convergence criteria for the Newton-Raphson method is given as:

$$max(\frac{\|\Delta\phi_i\|_2}{\|\phi_n\|_2}, \frac{\|\Delta u_i\|_2}{\|u_n\|_2}) \leq \epsilon = 1\times 10^{-4} \tag{64}$$

where $\Delta u_i$ and $\Delta \phi_i$ are the iterative changes in the displacement and phase-field, respectively, obtained from Eq. (54) at $i^{th}$ iterator of $(n+1)^{th}$ loading step. $\phi_n$ and $u_n$ are the phase-field and displacement values from the $n^{th}$ load step, respectively. Both second- and fourth-order phase-field theories are considered, including problems with and without the initial cracks, curved crack paths, crack coalescence and crack propagation through holes. In all of the analyses, full $(p+1)\times(q+1)$ Gauss quadrature is used to approximate the integrals over each element, where $p$ and $q$ are the orders of basis functions corresponding to the ξ- and η-direction, respectively.

## 4.1. Single edge notched under mode-I loading

In the first problem, a square plate with a side length of 1mm has a single edge notched on the left side with a length of 0.5mm, as illustrated in Figure 7a. The plate is subjected to tensile loading by applying a vertical displacement on the top edge and fixing the bottom edge. The material parameters used are the same as Miehe [48], include a shear modulus μ = 80.77 kN/mm$^2$, Lamé's parameter λ = 121.15 kN/mm$^2$ and critical fracture energy density $\mathcal{G}_C$ = 0.0027 kN/mm. In closely model a sharped crack, a length-scale parameter of $l_0$ = 0.0075 mm was chosen. Both second- and the fourth-order theories were used to analyse the problem. The incremental monotonic top edge displacement was $\Delta u = 1 \times 10^{-4}$ mm for the second-order model in the first 50 loading steps. In subsequent loading steps, the displacement increment was set to $\Delta u = 1 \times 10^{-6}$ mm. For the fourth-order model, displacement increments of $\Delta u = 1 \times 10^{-4}$ mm were applied in the first 40 loading steps and $\Delta u = 1 \times 10^{-6}$ mm in remaining loading steps.

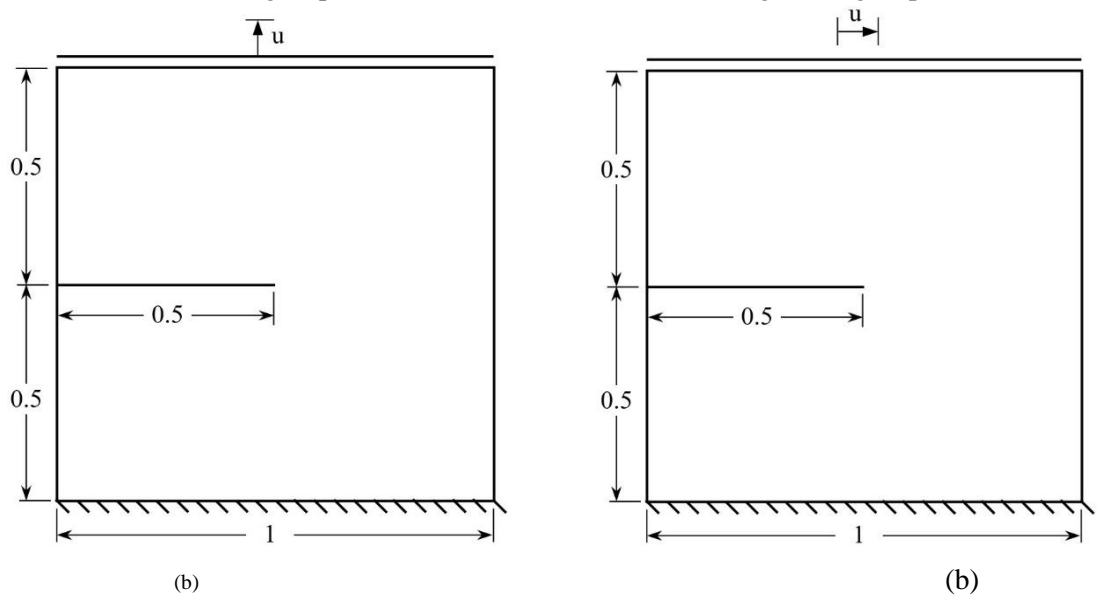

Figure 7. Boundary conditions and geometry of a single edge notched specimen under (a) mode-I and (b) mode-II loading.

In a crack propagation problem using phase-field model, the size of the mesh suggested by Miehe [11, 48] was very small to obtain an accurate solution when recovering the crack length. Here, by using IGA, a multipatch approach was used to build the model for this problem. The model used eight patches, which are illustrated in Figure 8a. In order to describe a strong discontinuity of the initial crack path, there was an interaction of the three variables (two displacement variables, $u_x$ and $u_y$, and phase-field variable, $\phi$) of control points located on the interface edge between the second patch and the third patch. Figure 8b shows the control point set in case of the coarsest mesh. From this coarsest mesh, patches 2, 3, 6 and 7 were refined to generate different meshes. Figure 9 displays a refined mesh with effective size level of $h = l_0/2$. The maximized element mesh size was proposed to be one half of the length-scale by Miehe [11] in order to aid the accuracy of the results. However, the ratio of the effective element size to the length-scale was chosen at approximately 7.5:1 in some practical problems [11]. The second-order phase-field theory is used in this part to compare with the proposed approach with Miehe's [48] solutions. It should be remarked that the Q4 element in the finite element method is a particular case of IGA using first-order basis functions. Here, a first-order of B-spline element is considered with multiple sizes located in the predicted crack propagation zone, for instance, $h = l_0/2$, $l_0/4$, $l_0/6$, in order to estimate the accuracy of the proposed method. Figure 10 shows that the finer mesh, the more accurate the results. The first-order mesh with the effective size $h = l_0/6$, which is assumed as a converged solution showed closely matching to the results published by Miehe [48]. However, the analysis involves a large number of DOFs (179364-DOFs), which is a result of a large memory requirement.

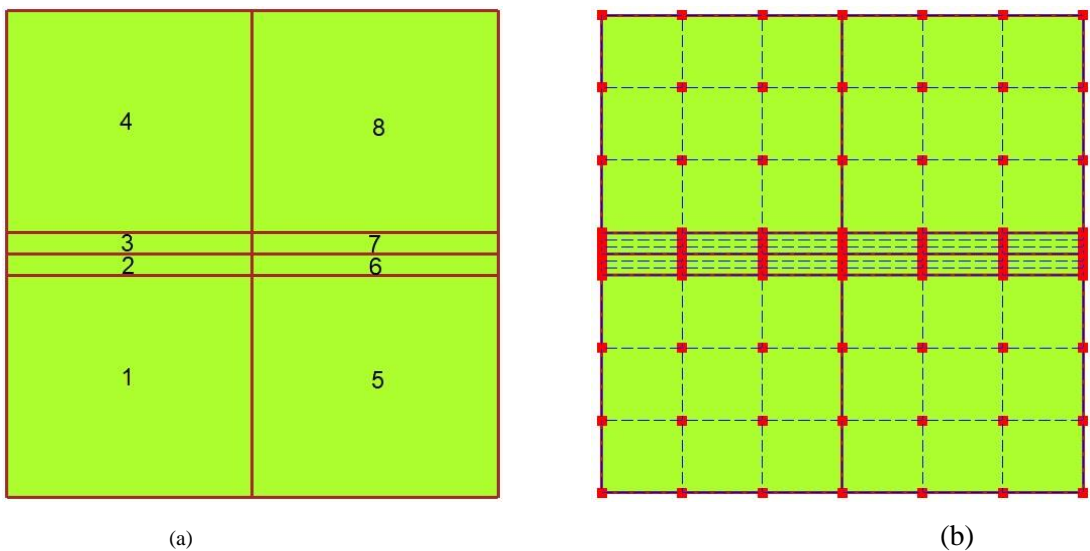

(a)                                            (b)

Figure 8. Multi-patch of single edge notched problem under mode-I loading: (a) patch definition and (b) the cubic NURBS control points for the coarsest mesh.

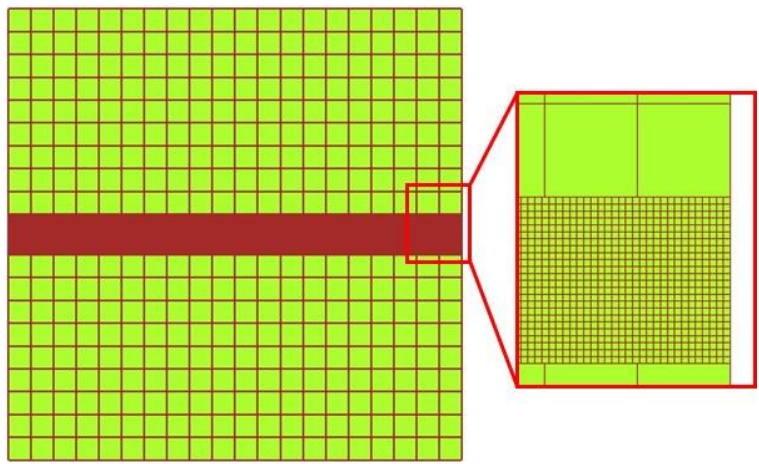

Figure 9. A refinement mesh of single edge notched problem under mode-I loading with the effective size of $h = l_0/2$.

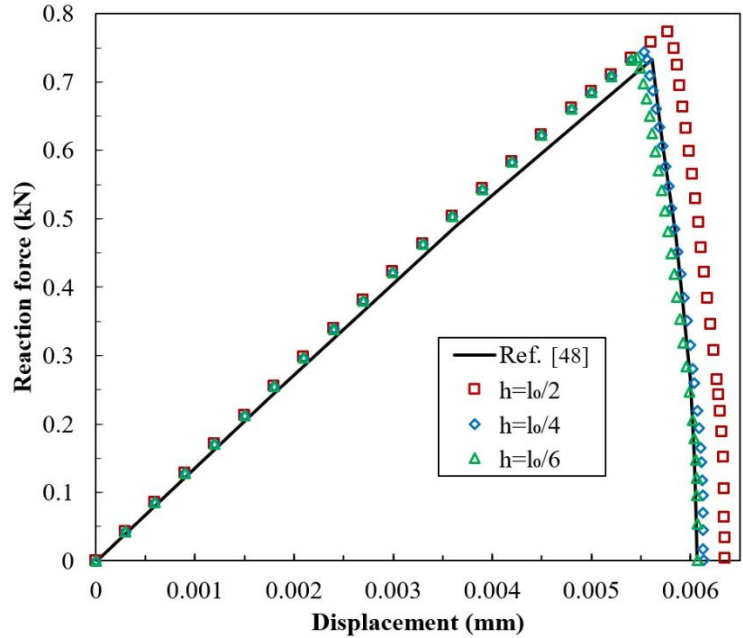

Figure 10. Reaction force versus displacement for various size of mesh of first-order B-spline elements.

In order to demonstrate the advantages of IGA in comparison with the traditional FEM, the results, higher-order B-spline elements, including cubic ($p = 3$) and quartic ($p = 4$) B-spline elements, are also used to analyse this problem. Various size meshes of higher-order B-spline elements is considered to estimate the accuracy of the present method, for instance, $h = 2l_0, l_0, l_0/2$. The results of the several orders B-spline elements are illustrated in Figure 11, whilst Table 1 provides the number of DOFs and the computational time for the different meshes. The obtained results of the crack path are in good agreement with the published results from Miehe [48]. However, the number of DOFs and computational time should be noted. It is easy to realise that a higher-order element has a higher computational cost due to the increase in the number of degrees of freedom. However, the solutions from the higher-order element can be more accurate than lower-order one when the number of DOFs is approximately the same. Specifically, the cubic B-spline element with an effective size of $h = l_0/2$ is seemly a good choice because of the above reason. Furthermore, the different orders of B-spline elements depicted in Figure 12, are considered to point out the exact solution because they have converged as the same solution with the element size of $h = l_0/2$. In this case, it is clear to see that the higher-order elements provide more accurate solutions than the lower-order one. Although the quartic B-spline element may gain a more accurate solution, it takes much more computational time with the same size of the mesh for relatively little improvement in accuracy, as shown in Table 1. Therefore, cubic-order B-spline elements with the effective size of $h = l_0/2$ will be used for all subsequent numerical analyses in order to balance the computational cost and the accuracy of the solutions. This choice has a much lower the number of DOFs and run-time than the first-order elements with the effective size of $h = l_0/6$.

Table 1

Computational time and number of DOFs in the different cases of order B-spline elements.

| Order B-spline elements | Number of DOFs | Computational time (min) |
|---|---|---|
| linear IGA ($h = l_0/2$) | 22224 | 140 |
| linear IGA ($h = l_0/4$) | 81720 | 560 |
| linear IGA ($h = l_0/6$) | 179364 | 1074 |
| cubic IGA ($h = 2*l_0$) | 4464 | 96 |
| cubic IGA ($h = l_0$) | 9432 | 220 |
| cubic IGA ($h = l_0/2$) | 26352 | 654 |
| quartic IGA ($h = 2*l_0$) | 5292 | 192 |
| quartic IGA ($h = l_0$) | 10704 | 420 |
| quartic IGA ($h = l_0/2$) | 28488 | 1276 |

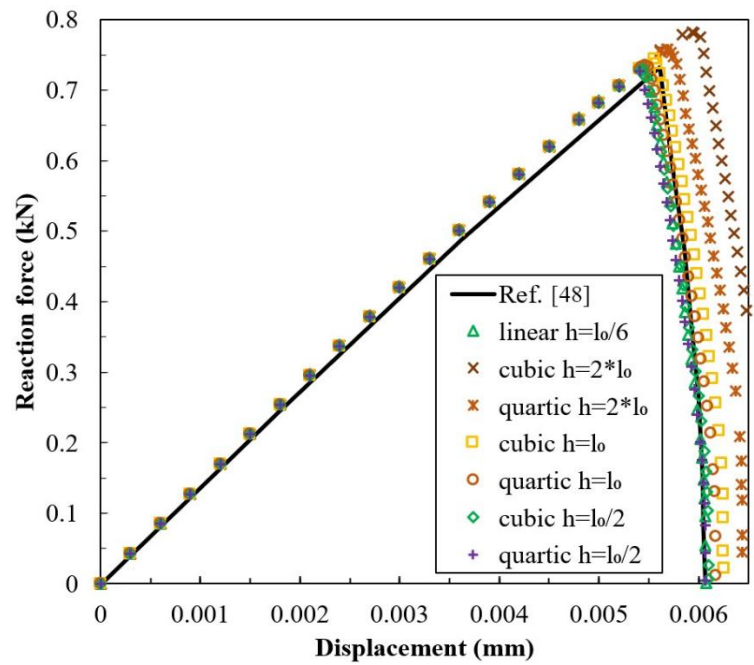

Figure 11. Reaction force versus displacement for various size of mesh of higher-order B-spline elements in IGA.

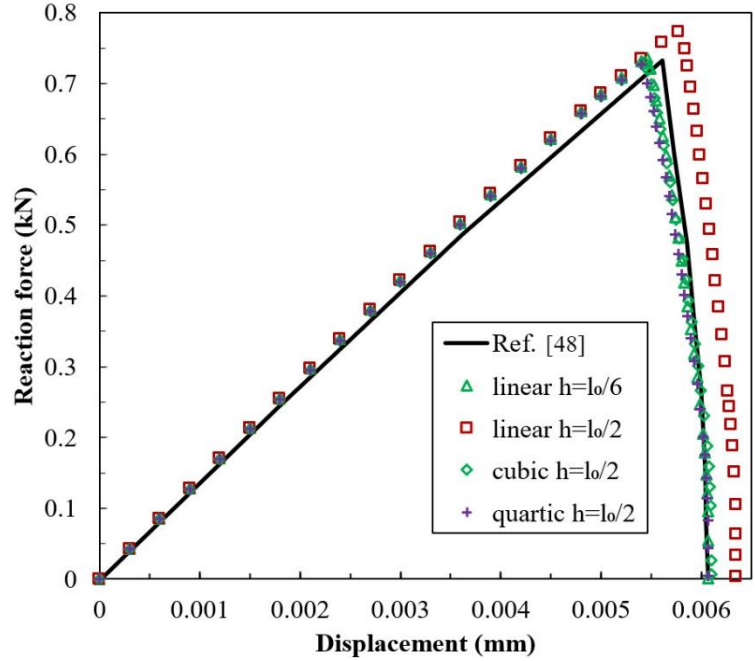

Figure 12. Reaction force versus displacement for various size of mesh of higher-order B-spline elements in IGA with the effective size of $h = l_0/2$.

The crack propagation for a single notched edge under mode-I loading is illustrated in Figure 13 by using cubic-order B-spline elements with the effective size of $h = l_0/2$ for two cases of phase-field theories (second- and fourth-order phase-field formulations). It is easy to realise that the latter illustrates more narrow crack than the former. Furthermore, for numerical solutions, Borden [35] proved that the fourth-order phase field formulation improves the rate of convergence, but it requires second derivatives of the underlying basis functions, which is satisfied by cubic-order B-splines. The difference between the two theories, depicted in Figure 14, has been shown as the same behaviour by Weinberg [55]. Last but not least, it should be noted that methods using enriched formulations, for instances, the extended finite element method (XFEM) [7], the extended isogeometric analysis (XIGA) [56] and the extended isogeometric boundary element method (XIBEM) [57], will face numerical difficulties in case of the cracks propagating to the boundary of the domain. In contrast, this situation can be studied easily by using the phase-field model, as shown in Figure 13b and Figure 13d.

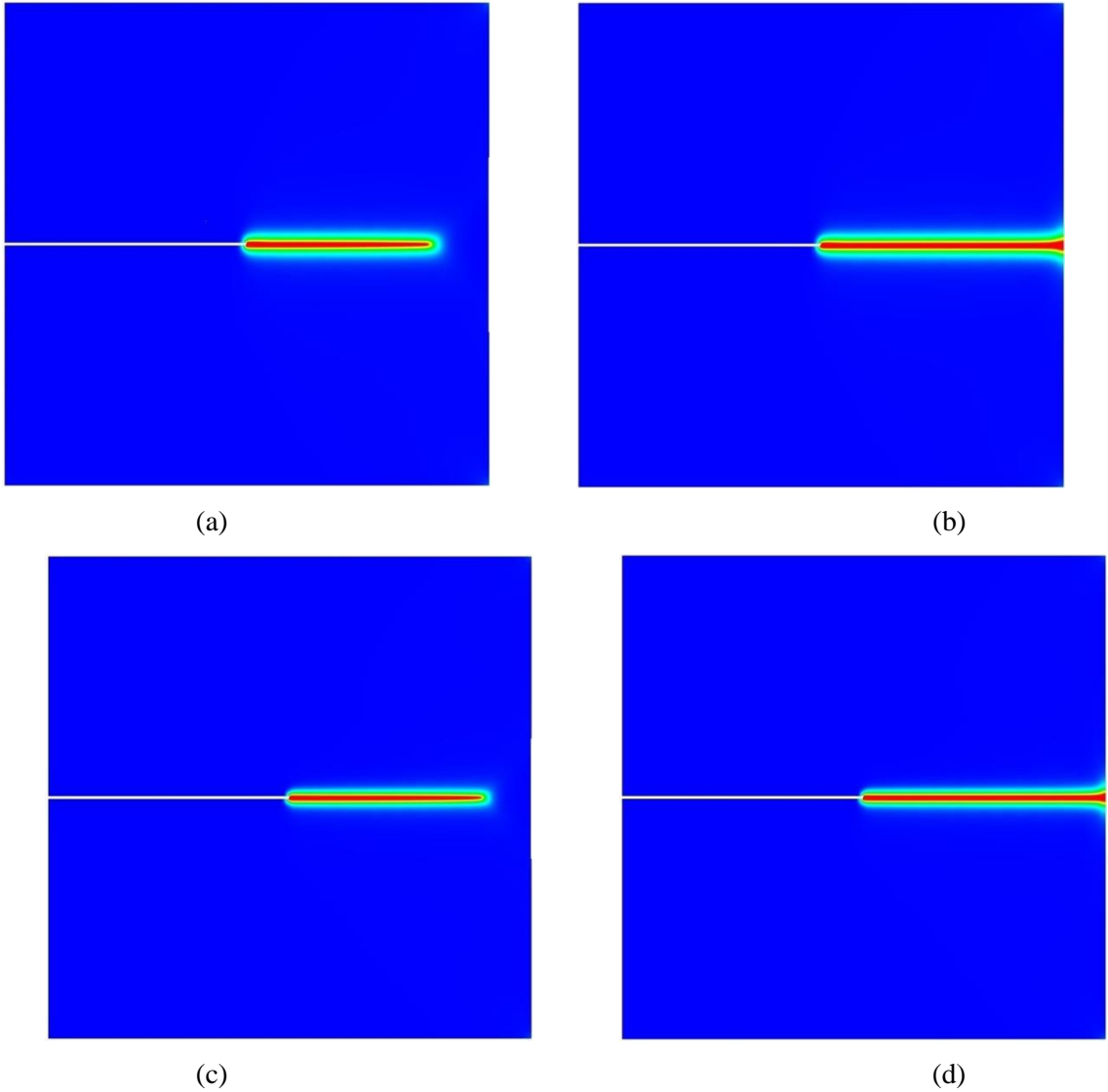

Figure 13. Crack propagation for single edge notched under mode-I loading with length-scale $l_0 =$ 0.0075 mm for (a) near fully separated plate, (b) fully separated plate in the case of second-order theory and (c) near fully separated plate, (d) fully separated plate in the case of fourth-order theory.

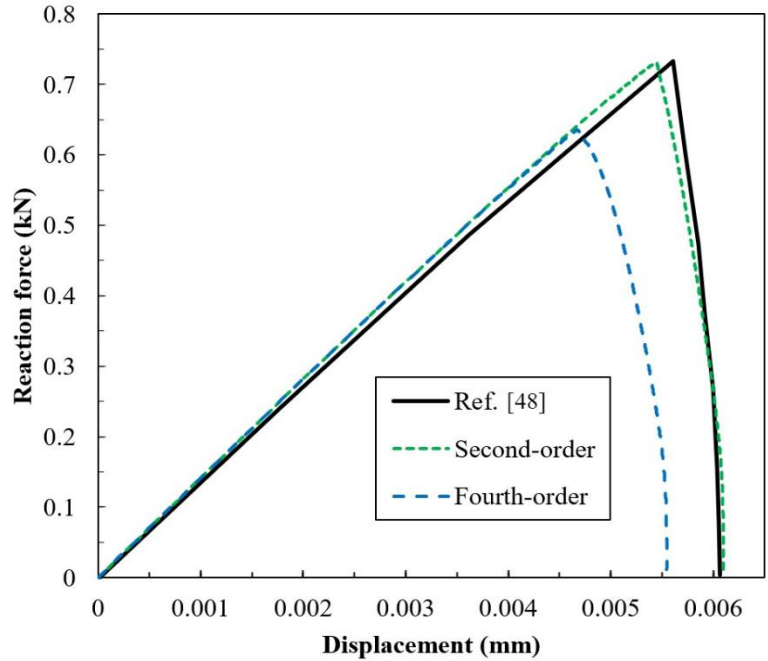

Figure 14. Reaction force versus displacement of single edge notched problem under mode-I loading in two cases of phase-field theories.

*4.2.    Single edge notched under mode-II loading*

The second example is a single edge notched unit square under pure shear loading – the boundary conditions are depicted in Figure 7b. According to Ambati [12], there are two formulations, hybrid [12] and anisotropic [11] formulations, that can be used to distinguish the positive and negative components of the strain energies under tension and compression. Here, the anisotropic formulation in Eq. (22) is used for all the following problems. Moreover, the parameters of material properties and the length scale are chosen the same as those used in Section 4.1. The coarsest mesh is illustrated in Figure 15, whereas Figure 16 displays a refinement mesh covering the predicted crack propagation zone. The model is divided into 16 patches in which seven patches are chosen to be refined (patches of 5, 6, 7, 9, 10, 11 and 13). As in the previous example, the VUKIMS coupling algorithm is used to locally refine the mesh and decrease the computational cost, including computational time and the required memory.

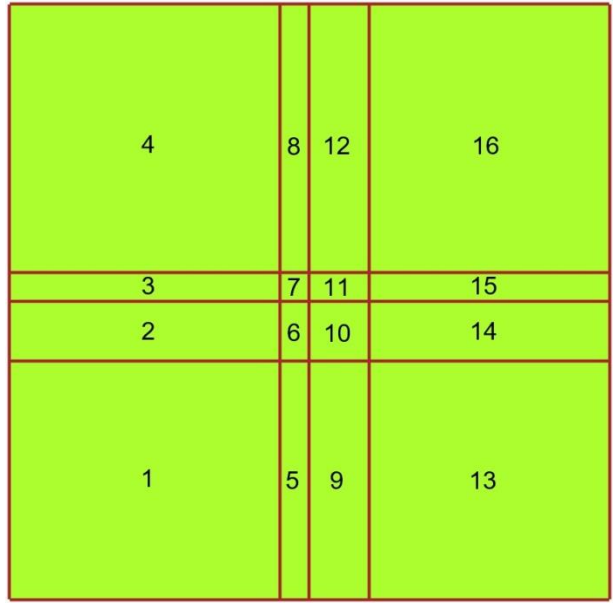

Figure 15. The coarsest mesh and patch numbering of single edge notched under mode-II loading.

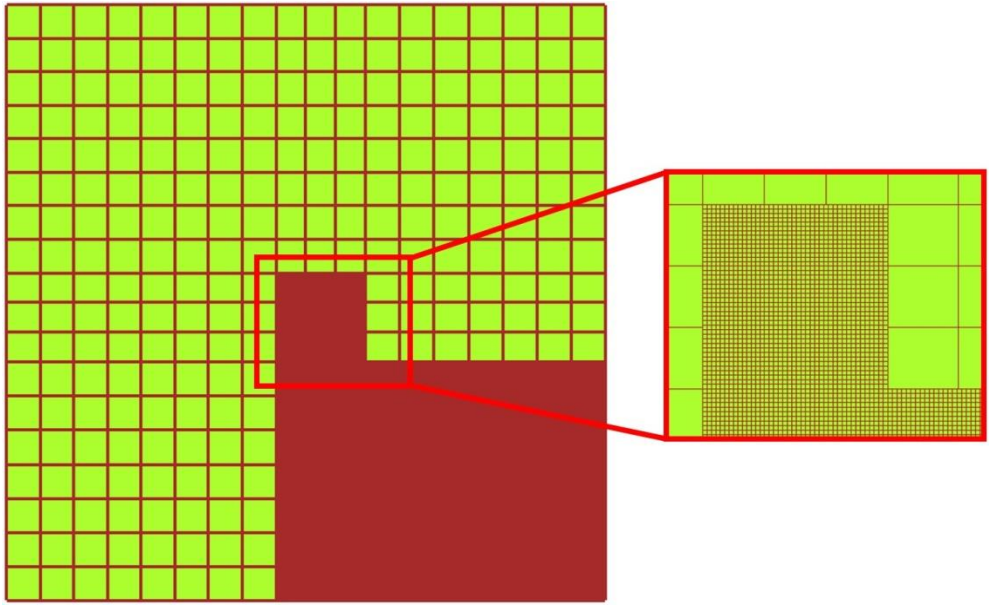

Figure 16. A refinement mesh of single edge notched under mode-II loading.

Based on the findings from the previous section, an effective element size is chosen as half of the length-scale value, and cubic B-spline elements are used for this example. For both the second- and fourth-order theories, the top edge is displaced by a monotonic horizontal displacement increment of $\Delta u = 1 \times 10^{-4}$mm for the first 80 loading steps and the increments are chosen as $\Delta u = 1 \times 10^{-6}$mm for subsequent loading steps. Both order phase-field results which are illustrated in Figure 17 are a well-matched solution in comparison with the results from [29, 48]. It should be noted that the proportion of the length scale parameter to the effective element size was approximately 3.75:1 in [48], and 1.88:1 in [29]. The former element size is smaller than the current elements, while the later is approximately equal. As a result, the result of the second-order solution is close to the result from [29]. Meanwhile, the result of the fourth-order solution is in good agreement with the result from [48]. Figure 17 reveals that the fourth-order solution is not only more accurate one but also predicts a more narrow crack than the second-order formulation, as shown in Figure 18. In addition, the topology of the crack path using the phase-field model is a smoother curve than these enriched formulations [7, 56, 57] which often use a multiple-line approximation to represent a crack propagated path.

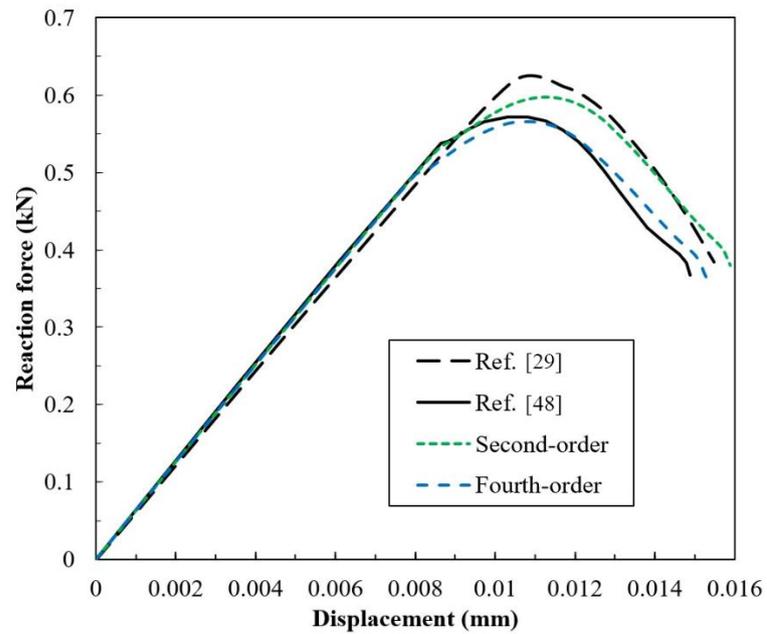

Figure 17. Reaction force versus displacement for a single edge notched problem under mode-II loading in two cases of phase-field theories.

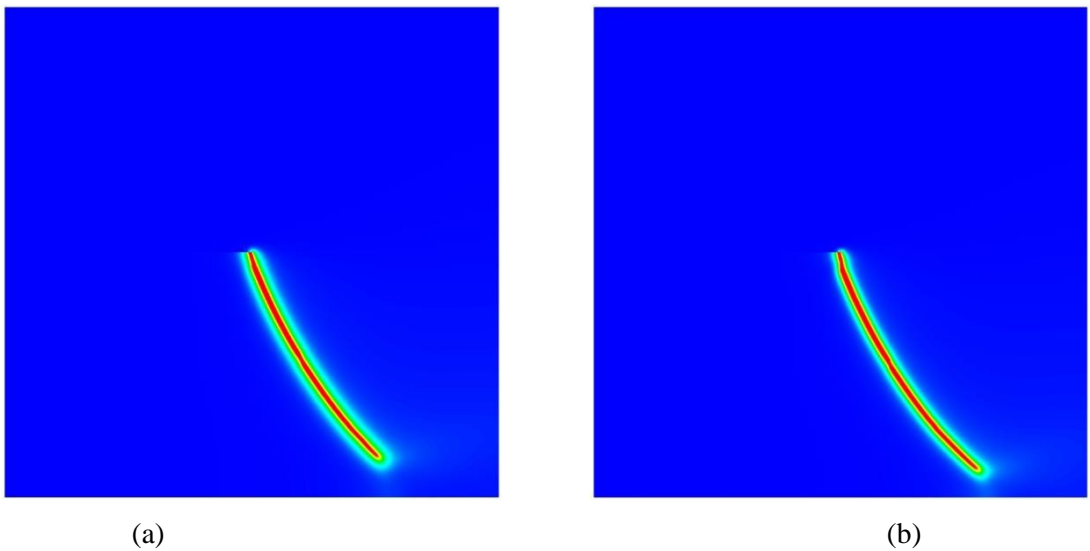

(a)                                                (b)

Figure 18. A crack propagation for single edge notched problem under pure shear loading with length-scale $l_0 = 0.0075$mm for (a) the second- and (b) the fourth-order phase-field theories.

*4.3. Symmetric three-point bending problem*

In this section, an asymmetric three-point bending example which is a well-known benchmark problem is considered. The material properties were chosen as follows: shear modulus $\mu = 8$ kN/mm$^2$, Lamé's parameter $\lambda = 12$ kN/mm$^2$, critical fracture energy density $\mathcal{G}_c = 0.0005$ kN/mm and a length-scale parameter of $l_0 = 0.03$ mm. The boundary conditions and geometry of this example are shown in Figure 19. In this problem, the model illustrated in Figure 20a is divided into four patches, in which the second and third patches of the model are selected to be refined in the expected crack propagation zone using the VUKIMS algorithm. The locally refined mesh is shown in Figure 20b. The effective element size is half of the length-scale number with cubic-order B-spline elements. Vertical displacement is applied to the top middle point of the beam with increments of $\Delta u = 1\times10^{-3}$ mm in the first 42 loading steps and $\Delta u = 1 \times 10^{-5}$ mm in remaining loading steps for both phase-field orders. The obtained solutions are illustrated in Figure 21 and are in good agreement with the solution published by Aldakheel [58]. Due to the symmetry of the geometry and boundary conditions, the crack propagation pattern displayed in Figure 22 is a straight, vertical path between the notch and the applied displacement. This path has been confirmed from some literature [12, 48, 58, 59].

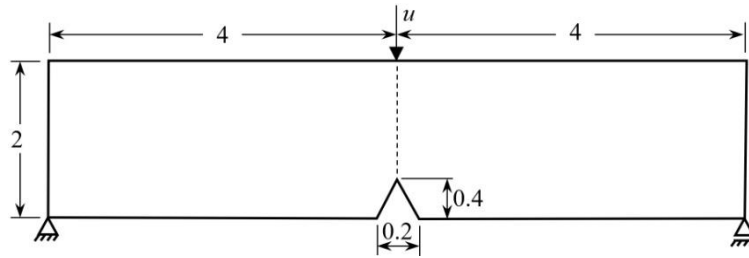

Figure 19. Symmetric three point bending specimen: boundary conditions and geometry.

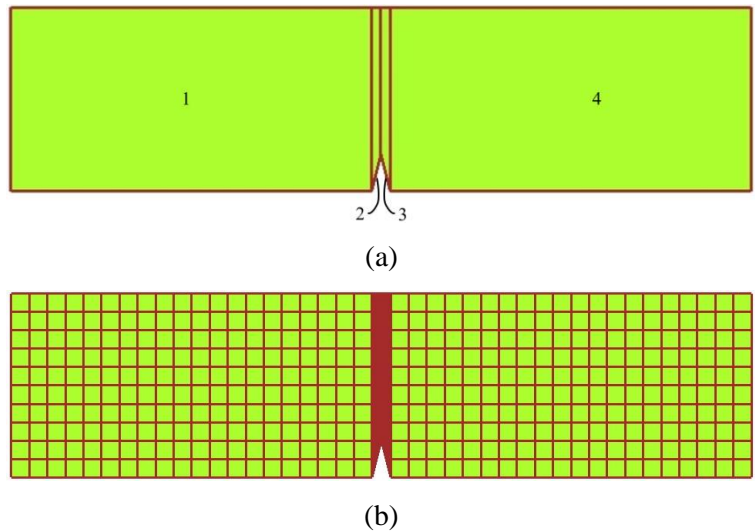

Figure 20. Three point bending: (a) coarsest mesh and patch numbering and (b) refined mesh.

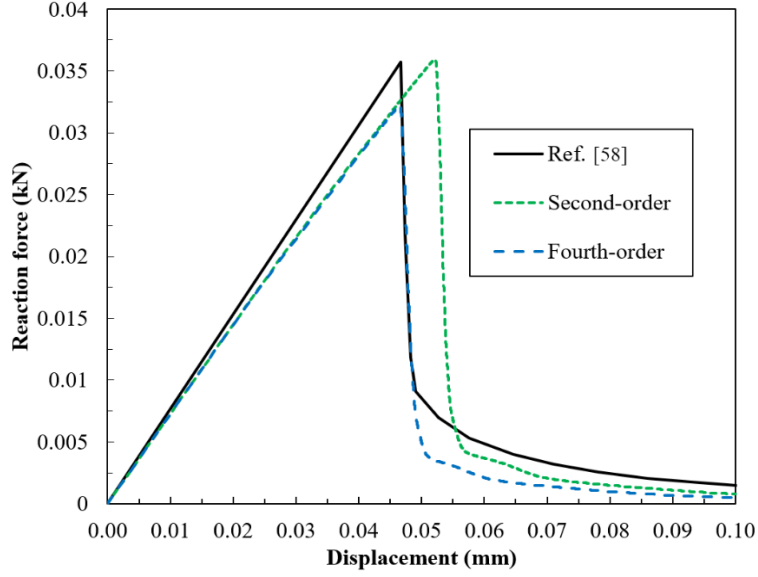

Figure 21. Reaction force versus displacement curves of symmetric three-point bending in two cases of phase-field theories.

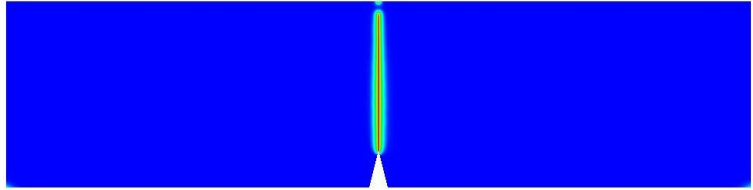

Figure 22. A crack propagation for symmetric three-point bending with the fourth-order formulation of the phase-field model at displacement $u = 0.1$ mm.

*4.4. Asymmetric double notched tensile specimen*

In this section, an asymmetric double notched of a rectangular plate is loaded under tension. The technique of pre-existing cracks is proposed by Borden et al. [36] in order to describe a simulation of arbitrarily multi-cracks. This method requires the initial strain-history field to be determined at each Gauss-point which located nearby the initial crack(s). The strain-history value at Gauss-point $i^{th}$ can be defined as

$$\mathcal{H}_i^+ = \begin{cases} \dfrac{B\mathcal{G}_C}{2l_0}\left(1 - \dfrac{2d_i}{l_0}\right) & d_i \leq \dfrac{l_0}{2} \\ 0 & d_i > \dfrac{l_0}{2} \end{cases} \qquad (65)$$

where $B$ is a scalar defined by $B = \dfrac{c}{1-c}$, $c$ should be approximately 1 (here taken to be $c = 0.9999$), $d_i$ is the closest distance from Gauss-point $i^{th}$ to the initial crack(s). Eq. (65) can be used to define multiple arbitrary cracks easily. If the initial cracks are nearly intersection or intersection each other, the strain-history field will be the maximum value of the strain-history values computed crack by crack. The boundary conditions and geometry of the analysed problem are shown in Figure 23. The problem is split into three patches, as shown in Figure 24a. In these patches, the middle patch is the predicted crack propagated zone and is refined, as shown in Figure 24b. The material properties are as follows: Young's modulus $E = 210$ kN/mm$^2$, Poisson's ratio $\nu = 0.3$, a length-scale parameter of $l_0 = 0.2$ mm and critical fracture energy density $\mathcal{G}_c = 0.0027$ kN/mm. The tensile loading is applied to the upper edge of the plate as via $\Delta u = 1 \times 10^{-3}$ mm in the first 32 loading steps and $\Delta u = 1 \times 10^{-6}$ mm in subsequent loading steps. Figure 25 plots the reaction force results of both phase-field model theories. The obtained crack path solution displayed in Figure 26 is in good agreement with many previous published solutions [52, 59, 60].

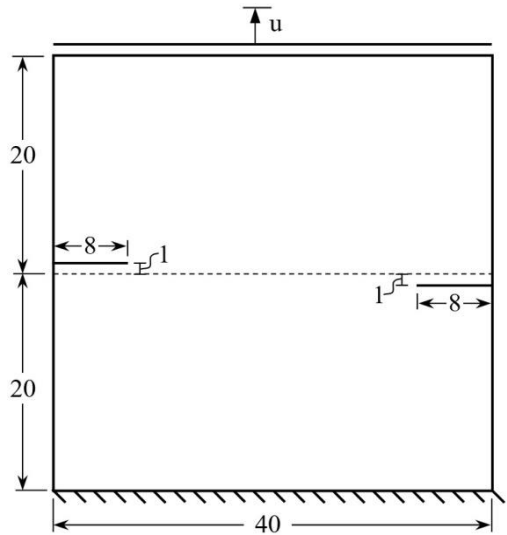

Figure 23. The asymmetric double notched tensile problem: boundary conditions and geometry [52].

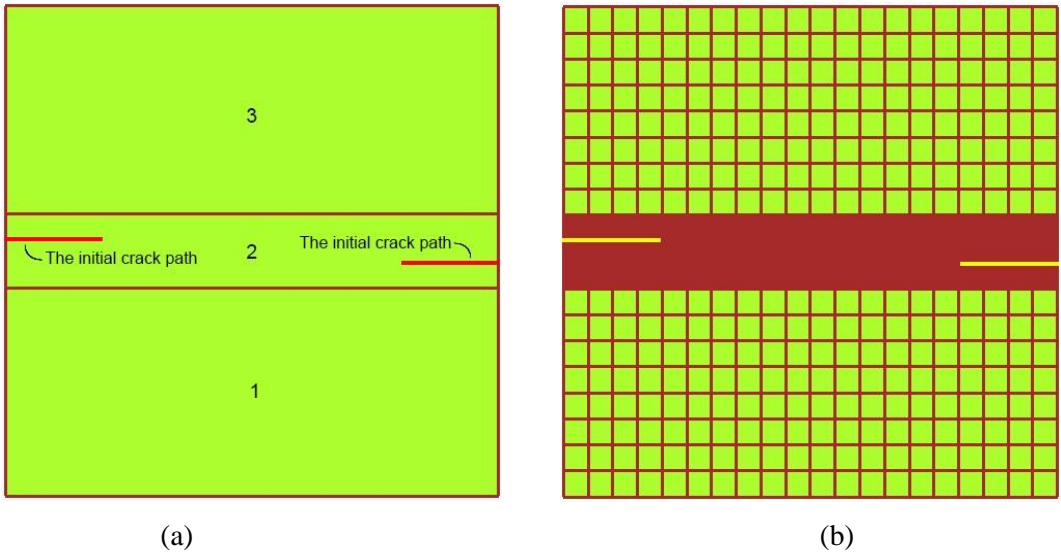

(a)                          (b)

Figure 24. Asymmetric double notched tensile specimen: (a) coarsest mesh and patch definition and (b) refined mesh.

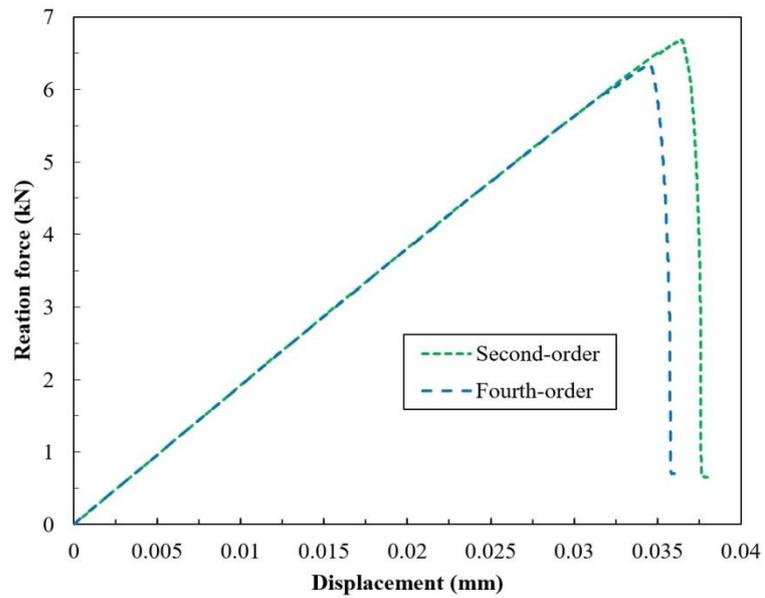

Figure 25. Reaction force versus displacement curves of the asymmetric double notched tensile problem in two cases of phase-field theories.

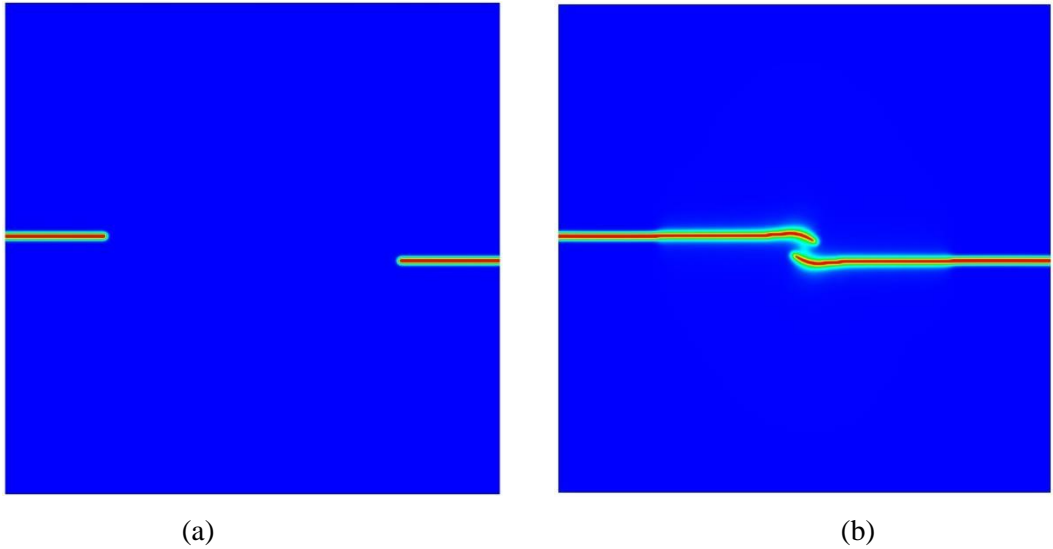

(a)                                                  (b)

Figure 26. Double-crack propagation for an asymmetric double notched tensile specimen corresponds the fourth-order phase-field theory at the (a) initial and (b) ending step.

*4.5. Notched plate with holes*

In order to illustrate the ability of the method to model curved and branching crack paths, a more complicated problem is considered in this section. The boundary conditions and dimensions of the problem are shown in Figure 27. Experimental data for this problem is available in Ambati et al. [12]. The specimen contains three holes which are conic geometries. NURBS elements are used to model these circles exactly (see Piegl [43]). In this case, the specimen is split into 43 patches and uses the cubic NURBS elements, as shown in Figure 29a. According to the predicted crack propagation zone from the experimental result revealed in Figure 28 from Ambati [12], six patches were refined (patch numbers of 4, 15, 23, 24, 32, 40), as illustrated in Figure 29b. The effective element size of these patches was set to be half of the length-scale number. The material properties were similar to Ambati [12]: shear modulus $\mu = 2.45$ kN/mm$^2$, Lamé's first parameter $\lambda = 1.94$ kN/mm$^2$, critical fracture energy density $\mathcal{G}_c = 0.00228$ kN/mm, as well as, a length-scale parameter of $l_0 = 0.3$ mm. The small lower hole is fixed while the small upper hole is displaced via a vertical displacement increment of $\Delta u = 1 \times 10^{-3}$ mm for all loading steps. Figure 30 illustrates the propagated crack path, which is solved by the current approach with the fourth-order phase-field model. Due to use phase-field model, the extraordinary benefit is observed that the second crack path propagated without any initial fracture, whereas this is not possible using enriched formulations, for instance, XFEM, XIGA and XIBEM. The current solutions are an excellent agreement with the results published in [12, 59] and the observed crack zone from four samples from Ambati's work in Figure 28b. Especially, the crack path in Figure 30d is the same in comparison with the experimentally crack patterns in Figure 28a. There is no significant difference between the results from the two phase-field theories, as shown in Figure 31.

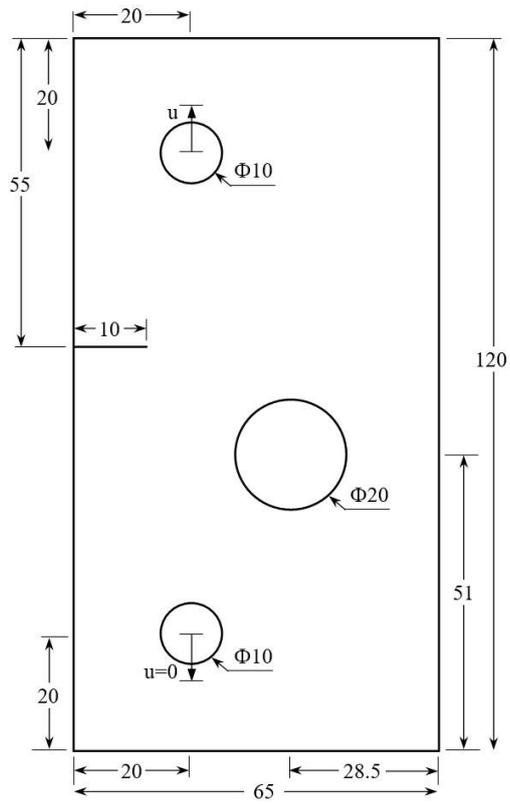

Figure 27. The notched plate with holes: boundary conditions and geometry.

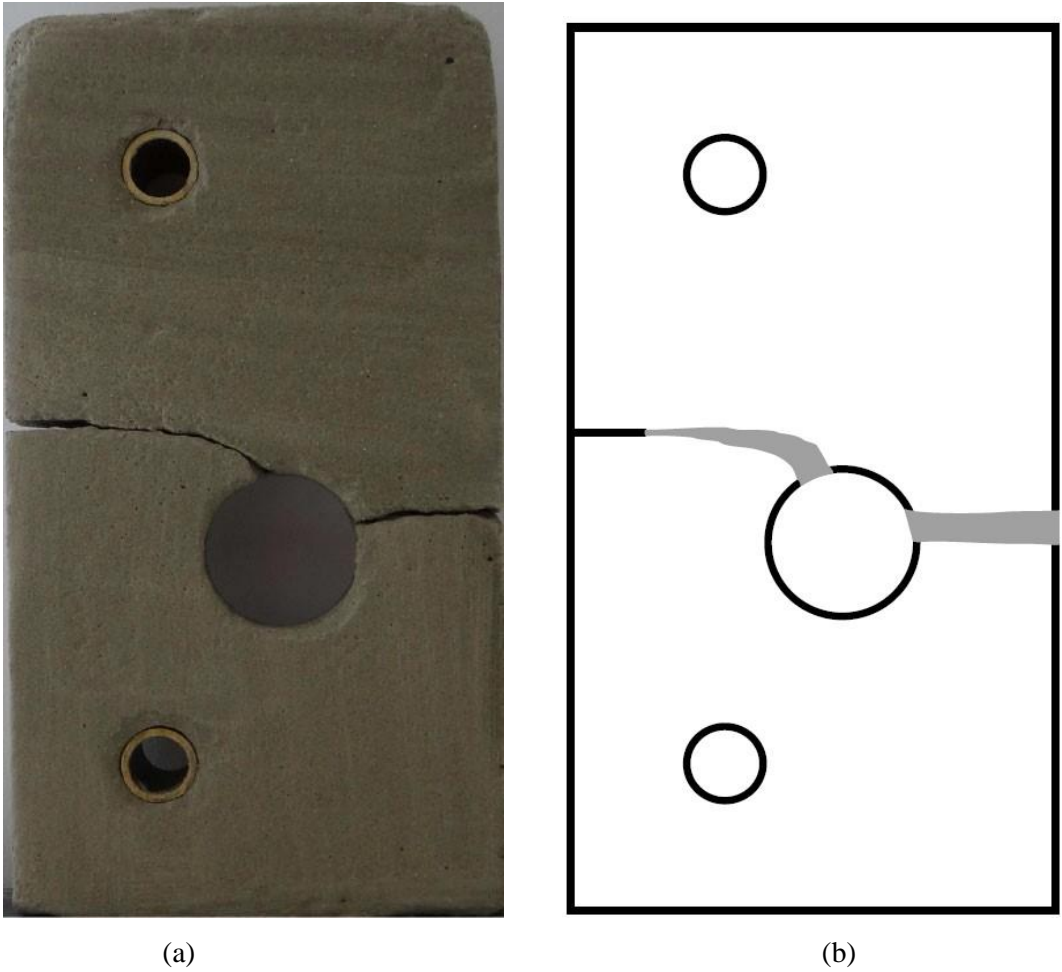

(a)                  (b)

Figure 28. The experimentally crack patterns of the notched plate with holes problem from Ref. [12]. (a) Fractured sample and (b) the observed crack path.

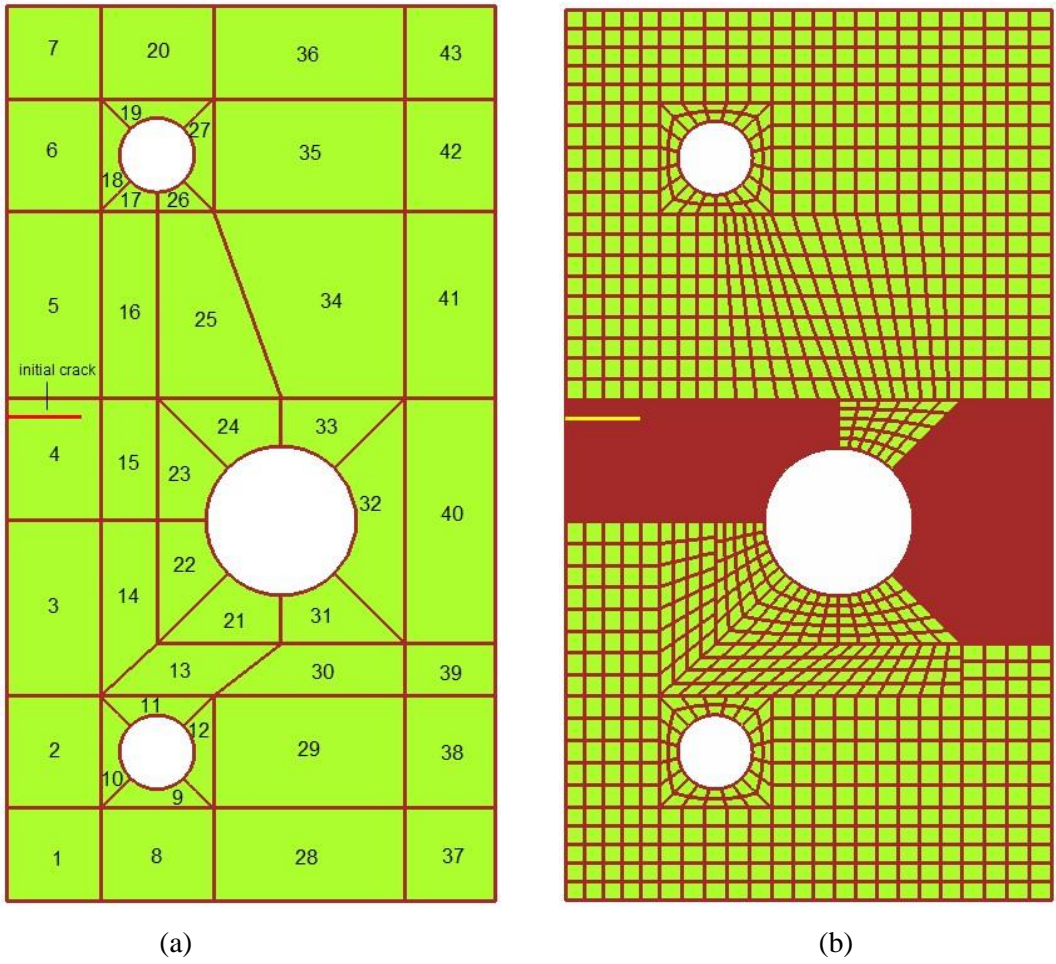

(a) (b)

Figure 29. The mesh of notched plate with holes example: (a) the coarsest mesh and patch numbering and (b) the refined mesh.

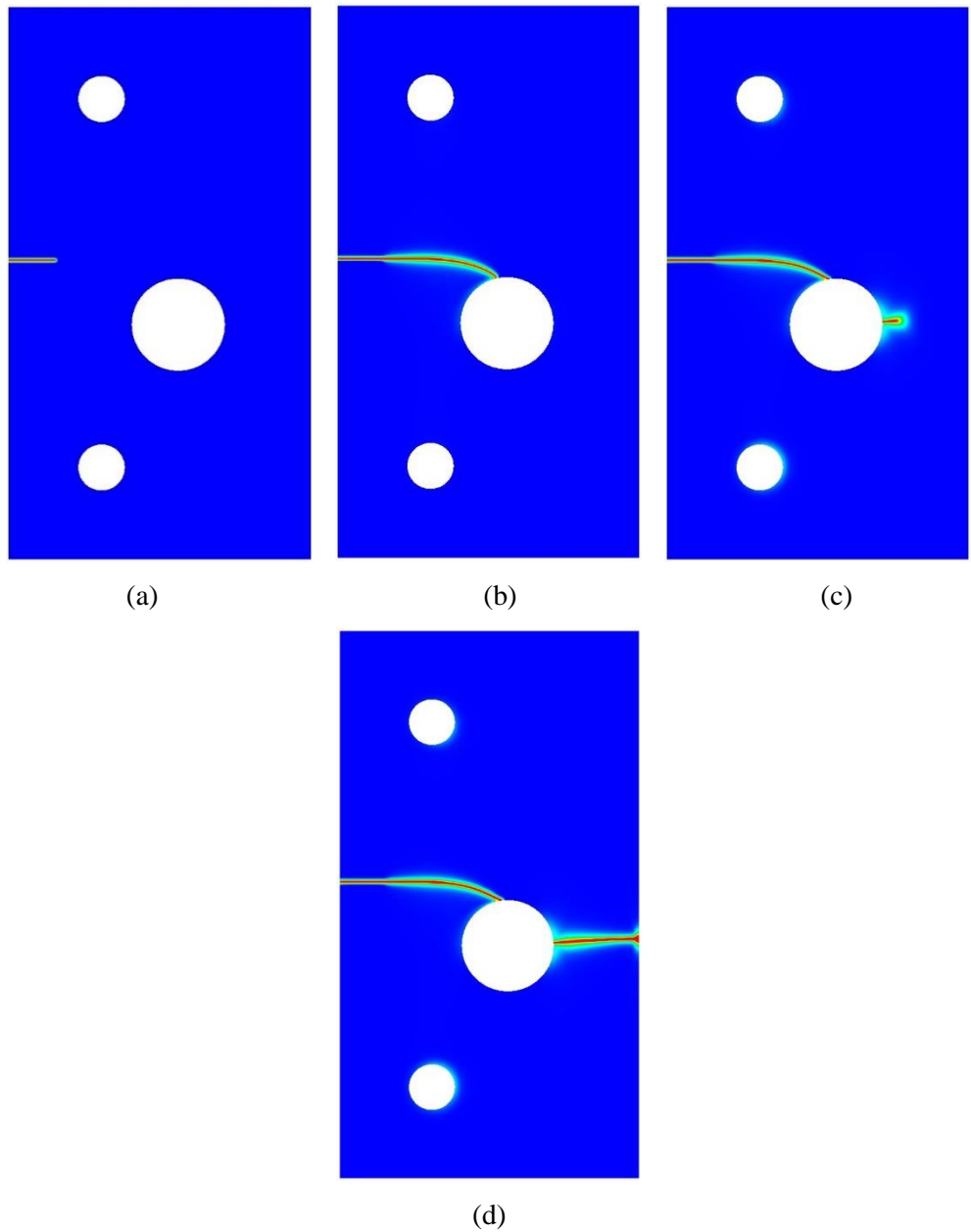

Figure 30. A notched plate example with holes: (a) the initial crack, (b) the propagated crack to the hole, (c) new appearance crack and (d) separated plate.

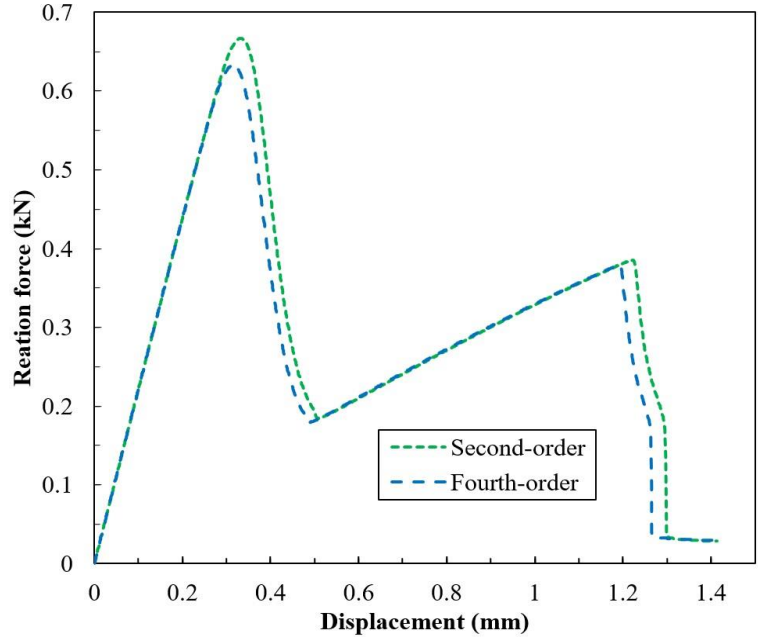

Figure 31. Reaction force versus displacement curves of the notched plate with hole example in two cases of phase-field theories.

## *4.6. Asymmetrically three-point bending*

In the last example, an asymmetrically three-point bending test, which is a well-known benchmark test in order to verify the solution of the proposed methods, is considered. For this study, parameters of *a* and *b* (see Fig. 32) are chosen as 6 mm and 1 mm, respectively. In addition, the material properties are chosen from [12] as follows: Lame's first parameter´ $\lambda = 12\text{kN/mm}^2$, shear modulus $\mu = 8\text{kN/mm}^2$, critical fracture energy density $G_C = 0.001\text{kN/mm}$ and a length-scale parameter of $l_0 = 0.043\text{mm}$. The boundary conditions and geometry of this problem are illustrated in Figure 32. Furthermore, the multi-patch model is shown in Figure 33a, while Figure 33b displays the locally refined mesh. Vertical displacement is applied to the middle point on the top edge of the plate via a displacement increment of $\Delta u = 1 \times 10^{-2}$mm in the first 12 loading steps and $\Delta u = 1 \times 10^{-5}$mm for the subsequent steps. The obtained results, which are displayed in Figure 34, are in excellent agreement with the solution from Patil [29]. In addition, Figure 35a and 35b show that the predicted fracture pattern is in good agreement with the experimentally observed fracture path from [61]. Additionally, the benefit of the anisotropic phase-field formulation is demonstrated by the crack which propagated through the second hole, while enriched formulations are unable to predict such behaviour without crack tip insertion.

Figure 32. An asymmetrically three-point bending test: boundary conditions and geometry.

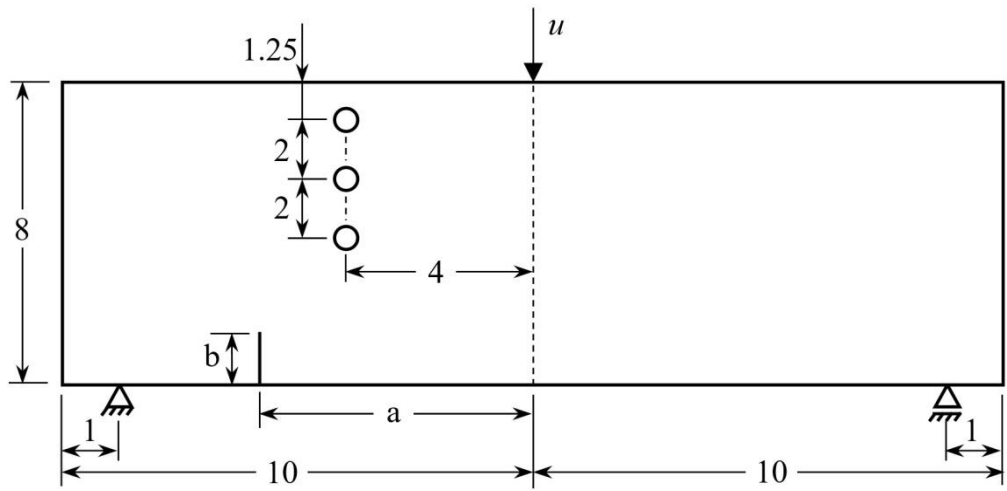

(a)

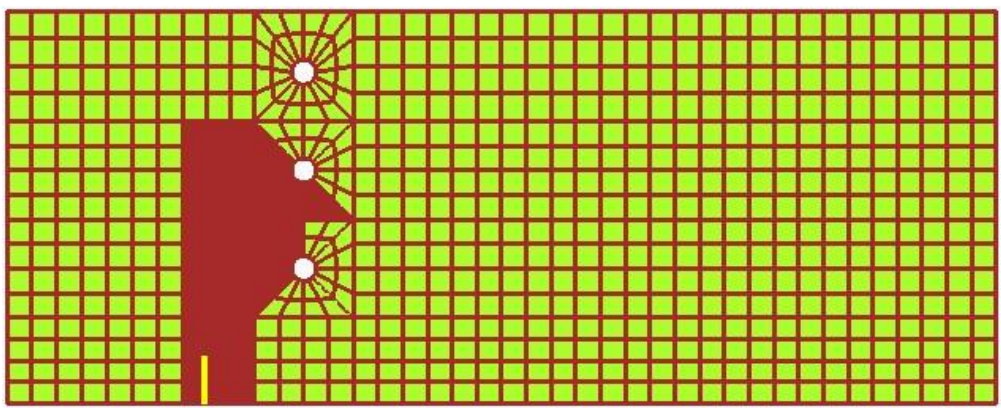

(b)

Figure 33. The mesh of asymmetric notched three-point bending problem: (a) the coarsest mesh and patch definition and (b) refined mesh.

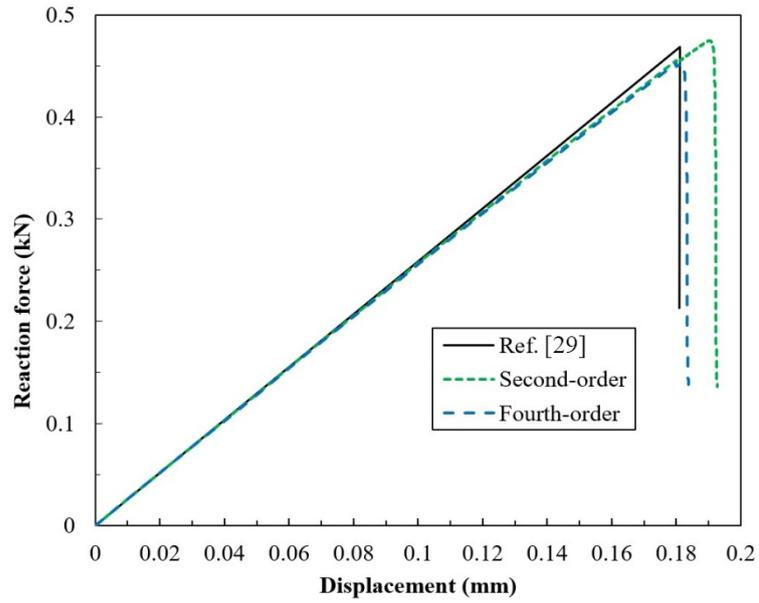

Figure 34. Reaction force versus displacement results of the asymmetrically three-point bending test in two cases of phase-field theories.

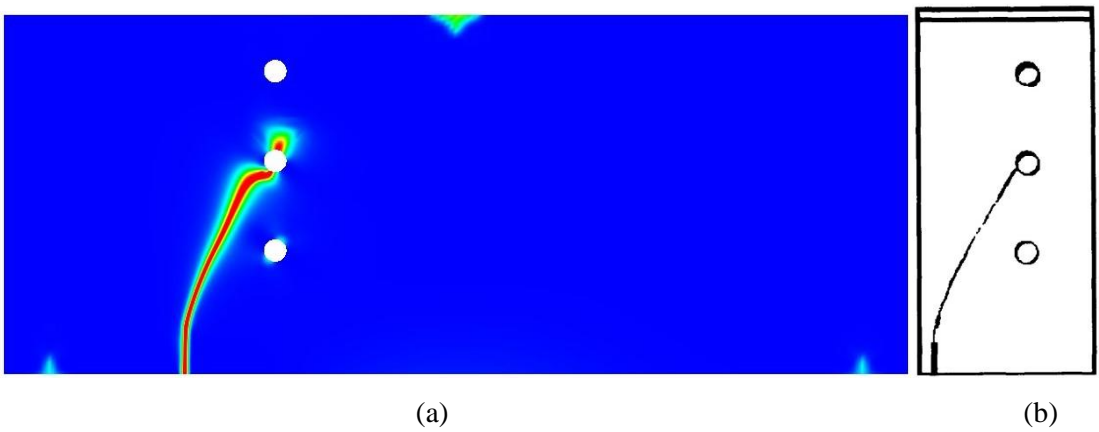

(a)  (b)

Figure 35. The asymmetrically three-point bending test: (a) crack path corresponds to the fourth-order formulation of the phase-field model, (b)experimentally observed crack pattern [61].

## 5. Conclusion

In this paper, we have proposed, validated and demonstrated an effective computational tool for brittle crack propagation based on a high-order phase-field model, combined with a non-conforming mesh of NURBS elements. In order to produce a locally refined mesh, a VUKIMS algorithm is proposed as an approach to deal with the computational cost of phase-field approximation. The algorithm allows for very small elements to be used to accurately model the damaged regions whilst allowing for coarser elements elsewhere. In addition, higher-order elements facilitate the use of higher-order phase-field theories which are proved to speed up the convergence rate for the numerical solutions. We found that the cubic NURBS elements are a good choice in terms of balancing the computational cost and the achieved accuracy, as demonstrated in Section 4.1. Several numerical examples have demonstrated the performance the approach in various crack geometries including with and without initial cracks, curved crack paths, crack coalescence and crack propagation through the holes. The combination of the phase-field approach and IGA is promising for the analysis of complicated problems in engineering practice.

**Acknowledgements**

The authors acknowledge the financial support of VLIR-UOS TEAM Project, VN2017TEA454A103, 'An innovative solution to protect Vietnamese coastal riverbanks from floods and erosion', funded by the Flemish Government and RISE-project BESTOFRAC (734370) is gratefully acknowledged.